\documentclass[11pt]{article}
\pdfoutput=1

\usepackage{euscript}
\usepackage{amssymb}
\usepackage{amsfonts}
\usepackage{amsbsy}
\usepackage{epsfig}
\usepackage{amsthm}
\usepackage{amscd}
\usepackage{amstext}
\usepackage{verbatim}
\usepackage{amsmath}
\usepackage{cancel}
\usepackage{capt-of}
\usepackage{empheq}
\usepackage{subfigure}
\usepackage{xcolor}

\usepackage[nosort]{cite}
\usepackage{bm}
\usepackage{authblk}
\usepackage[T1]{fontenc}

\usepackage[pdftex,linktoc=all]{hyperref}
\hypersetup{ 
colorlinks=true, 
linkcolor=black, 
citecolor=red, 
}

\def\ben{\begin{equation}}
\def\een{\end{equation}}

\let\a=\alpha    
   
\let\l=\lambda     
 \let\t=\tau

\let\w=\omega

\let\pa=\partial
\def\be{\begin{equation}}
\def\ee{\end{equation}}
\def\beq{\begin{equation}}
\def\eeq{\end{equation}}
\def\ba{\begin{array}}
\def\ea{\end{array}}

\def\dalemb#1#2{{\vbox{\hrule height .#2pt
       \hbox{\vrule width.#2pt height#1pt \kern#1pt
               \vrule width.#2pt}
       \hrule height.#2pt}}}

\newcommand{\bea}{\begin{eqnarray}}
\newcommand{\eea}{\end{eqnarray}}

\makeatletter
\newcommand*\bigcdot{\mathpalette\bigcdot@{.5}}
\newcommand*\bigcdot@[2]{\mathbin{\vcenter{\hbox{\scalebox{#2}{$\m@th#1\bullet$}}}}}
\makeatother

\renewcommand{\eqref}[1]{(\ref{#1})}

\title{Planckian Dissipation in Metals}
\author{Sean A. Hartnoll$^{\flat, \natural}$ and Andrew P. Mackenzie$^{\sharp,\dagger}$ \\
{\it $^\flat$ Department of Physics, Stanford University, Stanford, CA 94305-4060, USA} \\
\vspace{0.2cm}
{\it $^\natural$ Department of Applied Mathematics and Theoretical Physics,} \\
{\it University of Cambridge, Cambridge CB3 0WA, UK} \\
\vspace{0.2cm}
{\it $^\sharp$ Max Planck Institute for Chemical Physics of Solids,} \\
{\it N\"othnitzer Str. 40, 01187 Dresden, Germany} \\
\vspace{0.2cm}
{\it $^\dagger$ Scottish Universities Physics Alliance, School of Physics and Astronomy, University of St Andrews, St Andrews KY16 9SS, UK}
}
\date{}

\begin{document}

\maketitle

\begin{abstract}
   We review the appearance of the Planckian time $\tau_\text{Pl} = \hbar/(k_B T)$ in both conventional and unconventional metals. We give a pedagogical discussion of the various different timescales (quasiparticle, transport, many-body) that characterize metals, emphasizing conditions under which these times are the same or different. Throughout, we have attempted to clear up aspects of the problem that had been confusing us, in the hope that this helps the reader as well. We discuss the possibility of a Planckian bound on dissipation from both a quasiparticle and a many-body perspective. Planckian quasiparticles can arise naturally from a combination of inelastic scattering and mass renormalization. Many-body dynamics, on the other hand, is constrained by the basic time- and length- scales of local thermalization.
\end{abstract}

\newpage

\tableofcontents

\section{Planckian scattering from Peierls to the present}
\label{sec:intro}

\subsection{The Planckian timescale}
\label{sec:intro2}

The timescale
\be
\tau_\text{Pl} = \frac{\hbar}{k_B T} \,,
\ee
has long been known to control the electronic dynamics of the cuprate strange metal, as probed by optics \cite{PhysRevB.39.6571, PhysRevB.42.6342, Marel2003}, photoemission \cite{Valla2110} and, more recently, analysis of dc transport data \cite{Legros2019}. This timescale has also long been associated to quantum criticality \cite{PhysRevLett.60.1057, PhysRevB.39.2344} and known to bound the validity of a Boltzmann description of transport \cite{P2, peierls1996quantum}. It was first termed `Planckian' in Zaanen's \cite{Zaanen2004} analysis of an observed correlation between the superfluid stiffness and the resistivity at the superconducting transition temperature in several cuprates \cite{Homes2004}. The name `Planckian' references the necessarily quantum-mechanical origin of $\tau_\text{Pl}$ but also evokes the idea of a shortest possible timescale, by analogy to the `Planck' time in quantum gravity. This is the notion that suitably defined timescales $\tau$ should obey the `Planckian bound'
\be\label{eq:bound}
\tau \gtrsim \tau_\text{Pl} \,.
\ee
A bound of this form had previously been suggested 
in \S2.2 of \cite{subirbook}, with $\tau$ being the phase coherence time in quantum critical systems. We will refer to fast decay rates $1/\tau \gg 1/\tau_\text{Pl}$ as `super-Planckian' while slow decay rates $1/\tau \ll 1/\tau_\text{Pl}$ are `sub-Planckian'.

An important point becomes apparent when examining (\ref{eq:bound}): The finite quasiparticle lifetime due to elastic scattering from disorder at $T=0$, responsible for the residual resistivity of metals, necessarily violates the Planckian bound, because $1/\tau_\text{Pl} \to 0$ as $T\to0$. This super-Planckian
decay rate arises because the single-particle Bloch states are a broad superposition of the
single-particle energy eigenstates, due to explicitly broken lattice translation invariance.
This broadening is unrelated to many-body interactions, and suggests that a Planckian
bound should apply only to inelastic dynamics that is capable of redistributing energy between
particles and thermalizing the system. We return to this point in \S\ref{sec:notMIR}.

\begin{figure}[h]
    \centering
    \includegraphics[width=0.6\textwidth]{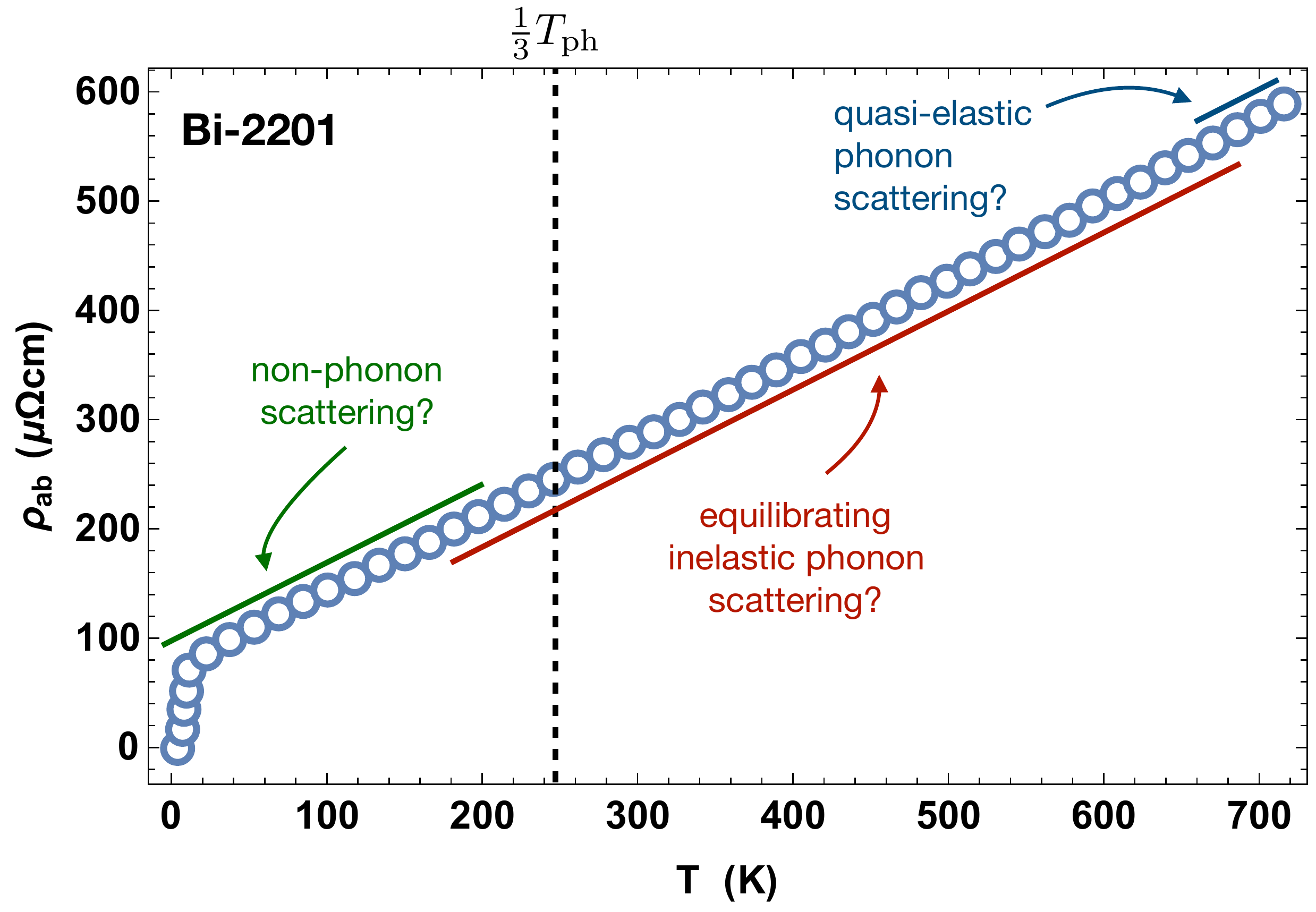}
    \caption{$T$-linear resistivity from $T_\text{c} \approx 7$ up to 700 K in Bi-2201, from \cite{PhysRevB.41.846}.
    ARPES measurements of the nodal electrons, which are believed to control transport, show a feature in their dispersion due to coupling to a known optical phonon in this material at around 63 meV \cite{PhysRevLett.100.227002}. In the plot we have indicated the corresponding temperature scale $\frac{1}{3} T_\text{ph}$ where $T$-linear scattering by this mode is expected to onset. Below this temperature the resistivity is not due to conventional electron-phonon scattering by this mode. At temperatures above $T_\text{ph}$, scattering by this phonon mode becomes increasingly elastic, as we discuss further in \S\ref{sec:elas} below. This plot aims to illustrate, schematically, that different mechanisms for $T$-linear resistivity, with the same slope, may plausibly be at work in a given unconventional metal.}
    \label{fig:martin}
\end{figure}

In the following sections we will describe two very different mechanisms leading to a Planckian electronic lifetime: quantum criticality and scattering by lattice vibrations. While quantum criticality has links with Planckian transport {\it below} a system-dependent characteristic temperature, Planckian transport due to electron-phonon scattering sets in {\it above} one. Even allowing for 
a low Bloch-Gr\"uneisen temperature \cite{PhysRevB.99.085105}, phonon scattering cannot explain observed $T$-linear scattering down to zero temperature. Therefore, it is likely that both mechanisms are present in unconventional metals \cite{Mousatov:2020auf}. See Fig.~\ref{fig:martin}
for an illustration of this likelihood.
If there are indeed different scattering mechanisms at work, this figure raises the question of how these mechanisms manage to seamlessly `pass the baton' without any feature or regime of additive scattering. As noted in \cite{Bruin804}, it is tempting to rationalise this observation by postulating the existence of a universality that transcends microsopic detail.  Such a postulate, however, raises at least as many questions as it answers.  In the remainder of this review we will further assess whether the evidence supports the notion of universal Planckian behavior, and discuss possible origins for it.

\subsection{Planckian scattering and quantum criticality}
\label{sec:qc}

It is natural to associate the Planckian time with quantum criticality, as emphasized in \cite{subirbook}.  Quantum critical systems exhibit an accumulation of low energy modes such that there is an emergent scaling symmetry at small frequencies $\omega$ under $\omega \to \lambda \, \omega$. Because temperature appears in the partition function as a weighting of states by their energy, low temperatures must also scale as $T \to \lambda \, T$. It follows that, in the absence of other scaling variables such as wavevector, dimensionless physical observables are scaling functions $F(\frac{\hbar \omega}{k_B T}) = F(\omega \tau_\text{Pl})$. If such a function does not have large or small dimensionless numbers in it, it necessarily varies over Planckian times. That is, $\tau_\text{Pl}$ is the characteristic timescale of a quantum critical system, below some characteristic energy scale where scale-invariance emerges. This expectation is confirmed by explicit computations in models of quantum critical magnets and superfluids \cite{subirbook}. Scaling of observables with $\omega/T$ has been reported in several measurements of strange metals, including photoemission \cite{Valla2110, Reber2019}, neutron scattering \cite{PhysRevB.46.14034, Aeppli1432,Schroder2000} and optical conductivity \cite{Marel2003, Prochaska285}. These results are suggestive of quantum criticality at work, as is the widespread conjunction of strange metallicity with quantum critical points \cite{sachdevkeimer}.

However, there are also some general obstacles that arise in attempting to deduce Planckian transport as a consequence of quantum criticality. In metals, low energy fermionic excitations are supported on a nontrivial locus in momentum space (the Fermi surface), while low energy excitations of the bosonic order parameter are typically either long-wavelength or supported close to a specific density wave ordering vector. This kinematic tension causes difficulties for the theoretical description of these systems and also, we will see shortly, weakens the direct imprint of criticality on transport. For recent discussions on theoretical approaches, including comparison to Monte Carlo numerics, see \cite{erezMC, erezPRX}. In the following paragraphs we discuss two elementary considerations for transport.

Firstly, long wavelength critical modes cause only small angle scattering of the (short wavelength) fermions, and this scattering does not degrade current efficiently.\footnote{Exceptions to this statement may occur if there are sharp corner-like features in the Fermi surface, as these allow small changes in momentum (i.e.~going across the corner) to give rise to large changes in current.} We can illustrate this phenomenon with a ferromagnetic quantum phase transition in three dimensions. While this critical point is preempted by a first order transition \cite{RevModPhys.88.025006}, the `pre-asymptotic' scaling regime is both under theoretical control, e.g.~\S 18 of \cite{subirbook}, and corroborated experimentally \cite{Smith2008}. The Landau-damped bosonic order parameter has $z=3$ dynamic scaling $\w \sim k^3$. Scattering from this order parameter is seen to lead to a `marginal' zero temperature fermionic self-energy $\Sigma \sim \omega \log \omega$. However, the resistivity is $\rho \sim T^{5/3}$. A factor of $T$ duly comes from the inverse quasiparticle lifetime, while a small-angle factor comes from the momentum exchanged between the fermion and the boson: $(\Delta k)^2 \sim \omega_\text{bos}^{2/3} \sim T^{2/3}$. Quantum criticality, therefore, does not automatically imply $T$-linear resistivity. The two dimensional marginal Fermi liquid of \cite{PhysRevLett.63.1996} has the same electronic self-energy, but evades the small-angle factor in the resistivity by postulating critical scattering
over a temperature-independent range of momenta. This then leads to $T$-linear transport. We will discuss Planckian transport in marginal Fermi liquids in more depth in \S\ref{sec:MFL} below. More general critical scenarios may similarly lead to $T$-linear transport at intermediate temperatures if the momentum transfer to the critical boson becomes large relative to the scales of the Fermi surface geometry, again evading small angle factors. This effect may underpin the $T$-linear resistivity reported in numerical studies of nematic quantum criticality \cite{doi:10.1073/pnas.1620651114}.

Secondly, critical density wave modes only efficiently scatter `hot spots' on the Fermi surface that are connected by the ordering vector. It is well-known that fermions at the hot spots will be short circuited in transport probes by longer lived and less heavy `cold' fermions that are not efficiently scattered by the critical mode \cite{PhysRevB.51.9253, PhysRevLett.82.4280}. Again, the resistivity is not necessarily $T$-linear. It may be possible for the cold fermions to become `lukewarm' \cite{PhysRevB.84.125115} through interactions with the hot spot excitations. In this way the influence of the quantum critical modes is indirectly felt around the whole Fermi surface. In particular, a sufficiently strong peak in the density of states at hot spots can induce marginal Fermi liquid behavior over the entire Fermi surface due to cold fermions scattering into the hot spots \cite{Mousatov2852}. Finite wavevector fluctuations have also been argued to influence the entire Fermi surface \cite{caprara2020dissipationdriven}.

The above considerations demonstrate that, while there are clear empirical and theoretical links between quantum criticality and Planckian timescales, quantum criticality alone is not sufficient to give Planckian transport.  Neither is it necessary, as we now discuss.

\subsection{Planckian scattering from phonons in conventional metals}
\label{sec:elph}

Although the modern interest in Planckian scattering was stimulated by the study of strange metals and often associated with quantum criticality, it has a far longer history than that, in a different setting. Peierls noted in 1934 that conventional metallic elements at room temperature host Planckian electrons \cite{P2, peierls1996quantum}. We reproduce his numbers in 
Table \ref{tab:P} below. Some decades later, Devillers identified Planckian scattering at room temperature in over twenty elements and compounds \cite{DEVILLERS19841019}. The appearance of the Planckian time in both conventional and unconventional metals was first emphasized in \cite{Bruin804}.

\begin{table}[h]
    \centering
    \begin{tabular}{|c||c|c|c|c|c|c|c|c|c|}
    \hline
        \text{element} & \text{Ag} & \text{Au} & \text{Cs} & \text{Cu} & \text{K} & \text{Li} & \text{Mg} & \text{Na} & \text{Rb} \\
        \hline
        $\tau_\text{Pl}/\tau$ & 0.51 & 0.76 & 0.95 & 0.64 & 0.51 & 1.7 & 0.95 & 0.70 & 0.76 \\ \hline
    \end{tabular}
    \caption{Planckian elements at room temperature, as estimated in 1934 by \cite{P2}.}
    \label{tab:P}
\end{table}

The Planckian lifetime in these cases arises due to scattering of electrons by classical phonons. A given phonon mode with frequency $\omega_k$ is classical, i.e.~macroscopically populated, at temperatures $k_B T \gtrsim \hbar \omega_k$. Typically, enough phonons are classical for the resistivity to become $T$-linear above some fraction of the Bloch-Gr\"uneisen temperature $T \gtrsim \frac{1}{3} T_\text{BG}$ (in metals with large Fermi surfaces $T_\text{BG} \approx T_D$, the Debye temperature). It was recognized very early on that this `high temperature' $T$-linear scattering rate is rooted in the equipartition theorem for classical atomic vibrations. See \S VII.1 of \cite{mott1958theory} for an early discussion that attributes this observation to a 1913 paper by Wien. Specifically, the strength of scattering of electrons by a thermally vibrating atom is
\be\label{eq:A}
|V|^2 \sim \left\langle (\Delta x)^2 \right\rangle \sim \frac{k_B T}{K} \,.
\ee
Here $\Delta x$ is the displacement from equilibrium and $K$ is the atomic spring constant. The growth of this strength of scattering is responsible for the increasing resistivity. In three dimensions (\ref{eq:A}) can be interpreted as a geometric cross section of the vibrating atom.

A more accurate calculation must account for the wave nature of electrons and phonons, leading to the well-known Bloch-Gr\"uneisen theory. From this perspective the $T$-linear scattering (\ref{eq:A}) originates from the large number of phonon quanta $n_\text{ph} \propto T$, corresponding to classical atomic vibrations. For a large Fermi surface with $k_F \sim 1/a$, with $a$ the lattice spacing, the inverse electronic lifetime is found to be of order
\be\label{eq:first}
\frac{1}{\tau} \sim \frac{D^2}{K a^2 \times E_F} \frac{k_B T}{\hbar} \,.
\ee
Here $D$ is the deformation potential, $K a^2$ is the atomic binding energy and $E_F \sim \hbar v^\star_F k_F$ is the (renormalized) Fermi energy. It is notable that (\ref{eq:first}) does not depend on the ion mass, and in particular remains finite if the ion mass is taken to infinity and all scattering becomes elastic.
Because all lattice energy scales are ultimately rooted in the electronic dynamics that holds the atoms together, one can crudely estimate (cf. \S6.6 of  \cite{peierls1996quantum}) that $D \sim  K a^2 \sim E_F$.
This leads to a Planckian inverse lifetime $\tau^{-1} \sim k_B T/\hbar$.

An important point here is that despite the short and $\hbar$-dependent lifetime $\tau \sim \tau_\text{Pl}$, the mean free path is long: $\ell = v^\star_F \tau \sim (a E_F)/\hbar \times \hbar/(k_B T) = a \times E_F/(k_B T) \gg a$. This is because the Fermi velocity is large and has a quantum mechanical origin in the Pauli exclusion principle. That is, the electrons undergo many collisions per unit time because they move quickly. This is a difference with many unconventional Planckian materials which have short mean free paths and slower velocities \cite{Bruin804}.

The above estimates suggest that parametrically `super-Planckian' scattering could in principle arise in this context if the three energy scales $D$, $K a^2$ and $E_F$ become very different. For example, the electron-phonon coupling might be sufficiently strong --- or there might be a complicated unit cell with many atoms or more generally a large scattering phase space --- that the effective value of $D$ is large. Alternatively, $E_F$ might be renormalized down to a small value. Indeed, in conventional superconductors, such as Pb or Nb, the electron-phonon coupling is an order of magnitude larger than in the metals listed in Table \ref{tab:P} (see e.g.~\cite{allen}). At the highest temperatures this fact directly translates into super-Planckian scattering of electrons. However, as we elaborate in \S\ref{sec:notMIR} and \S\ref{sec:elas} below, this high-temperature scattering is elastic (for the electrons) and therefore not subject to a Planckian bound. At lower temperatures, of order the Debye scale, the scattering becomes inelastic. To obtain the physical electronic lifetime at these temperatures it is important to account for the temperature-dependent electronic mass renormalization (see \S\ref{sec:single}). Due to this renormalization, while the slope of the $T$-linear resistivity is constant through the Debye scale, the physical scattering rate is not. As we explain in \S\ref{sec:phdeb}, the mass renormalization is large when the coupling is large, and tends to restore a Planckian scattering rate. Using the measured low-temperature renormalized mass in the Drude formula gives $1/\tau \approx 3/\tau_\text{Pl}$ in Pb and Nb \cite{Bruin804}.


\subsection{Planckian times from Drude analyses (and beyond)}
\label{sec:drude}

Although analysis of dc transport is far from the most direct way to access information about Planckian timescales, as will be discussed thoroughly in \S\ref{sec:transportime} below, the field at present is in the common situation in which more dc transport data exists for systems of interest than data from other probes, motivating its assembly and analysis. In this section we summarize the results of a body of work that has extracted a timescale $\tau$ using the Drude formula for the dc conductivity: $\sigma = n e^2 \tau/m_\star$.\footnote{The analyses that follow were criticized for using the renormalized mass to extract a timescale from the Drude formula \cite{RevModPhys.92.031001, Sadovskii2020, Sadovskii_2021}. We believe that this criticism is misplaced. 
It is a true statement that, with assumptions that are discussed extensively in \S\ref{sec:single} and \S\ref{sec:transportime} below, the conductivity can be expressed in terms of the bare mass and the imaginary part of the single-particle self-energy $\Sigma''$. This formal object may be an interesting quantity vis-\`a-vis Planckian bounds, but is different to the physical scattering rate that sets the timescale over which `things happen' to the particle. The physical single-particle lifetime is determined by the pole of the full fermion Green’s function, see (\ref{eq:Green2}) below. To extract this timescale one must use the renormalized mass in the Drude formula, as was correctly done in the analyses that follow.} As we explain in \S\ref{sec:drude2} below, this formula can and must be refined to appropriately average over the contribution of light and heavy carriers. The essential point, however, is that to estimate a timescale in this way one needs to know the ratio of the charge density $n$ to the effective mass $m_\star$.

Several classes of materials with $T$-linear resistivity were considered in \cite{Bruin804}, using masses and densities obtained from low temperature quantum oscillations. In addition to some conventional metals, discussed in the previous \S\ref{sec:elph}, these included the heavy fermion materials UPt$_3$,~CeCoIn$_5$ and CeRu$_2$Si$_2$, the ruthenate Sr$_3$Ru$_2$O$_7$, the pnictide BaFe$_2$(P$_{0.3}$As$_{0.7}$)$_2$ and the organic superconductor (TMTSF)$_2$PF$_6$. Over the range of temperatures considered, the resistivity has the form $\rho = \rho_o + A_1 T$. The constant offset $\rho_o$ is often small in these particular materials and is subtracted out before performing the Drude analysis. Such subtractions are commonly done in the analysis of experimental data and implicitly make a physical distinction between elastic and inelastic scattering. As was mentioned in \S\ref{sec:intro2} above, and will be discussed further in \S\ref{sec:notMIR} below, the Planckian limit is only expected to apply to inelastic scattering and so this subtraction is well-motivated.\footnote{A simple subtraction assumes that elastic and inelastic scattering processes are additive and that the temperature-independent term is indeed due to disorder. While the validity of this assumption is not clear a priori in unconventional metals, which may not admit a quasiparticle description, controlled disordering of cuprates has shown that Matthiesen's rule is obeyed in weakly disordered, $T$-linear transport regimes \cite{PhysRevB.39.11599, PhysRevB.51.15653,PhysRevLett.76.684, Rullier_Albenque_2000, Clayhold2010}.\label{foot:dis}} It was found that all of the materials considered yielded a Planckian timescale from this analysis, as we show in Table \ref{tab:B} below.
\begin{table}[h]
    \centering
    \begin{tabular}{|c||c|c|c|c|c|c|c|c|c|}
    \hline
         & \text{Sr327} & \text{Ba122} & \text{TMTSF} & \text{UPt$_3$} & \text{Ce115} & \text{Ce122}  \\
        \hline
        $\tau_\text{Pl}/\tau$ & 1.5 & 2.2 & 0.9 & 1.1 & 1.0 & 1.1  \\ \hline
    \end{tabular}
    \caption{Planckian unconventional metals, as estimated from a dc Drude analysis by \cite{Bruin804}. The precise compounds are given in the main text. That same analysis estimated the values in Cu, Ag, Au, Al, Pd to be comparable to those quoted in Table \ref{tab:P} while, as we noted in \S\ref{sec:elph}, the values for Nb and Pb are closer to $\tau_\text{Pl}/\tau \approx 3$.}
    \label{tab:B} 
\end{table}

Several of the materials considered in \cite{Bruin804} were `bad metals', with very short mean free paths at high temperature \cite{PhysRevLett.74.3253}. The ubiquity of the Planckian timescale among bad metals led to the suggestion that a Planckian bound might control the incoherent dynamics capable of surpassing the `Mott-Ioffe-Regel' limit, $k_F \ell \gtrsim 1$ \cite{Hartnoll:2014lpa}. The relation between the Mott-Ioffe-Regel and Planckian limits will be the topic of the following \S\ref{sec:notMIR}.

A further example of a widely studied high temperature bad metal with $T$-linear resistivity is VO$_2$ above the metal-insulator transition \cite{PhysRevB.48.4359}. The lifetime has been extracted both from the optical conductivity and from dc Drude analyses \cite{PhysRevB.74.205118, Qazilbash1750, Lee371}. In the absence of quantum oscillations, masses have been estimated from both thermopower and optical measurements while the density was estimated from both the Hall conductivity and the expected number of conduction electrons. While there is not perfect quantitative agreement between the estimates, and the Drude peak is not a simple Lorentzian, the analyses consistently suggest an inverse lifetime with $\tau_\text{Pl}/\tau$ of order $10$, making VO$_2$ potentially the most super-Planckian of analyzed unconventional metals.

In the opposite regime, of low temperatures, a Drude analysis of dc transport in several families of overdoped cuprates (Bi2212, Bi2201, LSCO, Nd-LSCO, PCCO and LCCO) was performed in \cite{Legros2019}. These materials all show resistivity $\rho = \rho_o + A_1 T$ as $T \to 0$, with a magnetic field suppressing superconductivity. The residual term $\rho_o$ is subtracted off, as described above. In most cases the carrier masses were estimated from measurements of the specific heat and densities were obtained from the expected Luttinger count. The slope of the $T$-linear resistivity per CuO$_2$ plane increases upon lowering the doping and is significantly larger for hole-doped compared to electron-doped materials. These same trends were seen to occur in the effective mass such that a Planckian timescale is obtained from the Drude formula in all cases, as we show in Table \ref{tab:L} below.
\begin{table}[h]
    \centering
    \begin{tabular}{|c||c|c|c|c|c|c|c|c|c|}
    \hline
         & \text{Bi2212} & \text{Bi2201} & \text{LSCO} & \text{Nd-LSCO} & \text{PCCO} & \text{LCCO}  \\
        \hline
        $\tau_\text{Pl}/\tau$ & 1.1 & 1.0 & 0.9 & 0.7 & 0.8 & 1.2  \\ \hline
    \end{tabular}
    \caption{Planckian overdoped cuprates, as estimated from a dc Drude analysis of the $T \to 0$ resistivity by \cite{Legros2019}. See Table 1 of \cite{Legros2019} for the corresponding doping for each material.}
    \label{tab:L} 
\end{table}
An interesting complication that arises in this analysis is that in some cases the effective mass is logarithmically temperature dependent over the range of interest. We discuss this fact further in \S \ref{sec:MFL} and \S \ref{sec:drude2} below. Finally, it was noted in \cite{Legros2019} that because the low temperature $T$-linear term in the resistivity vanishes continuously for sufficient overdoping \cite{Cooper603}, as a function of doping the $T$-linear lifetime fills out the bounded range $0 \leq \tau_\text{Pl}/\tau \lesssim 1$. Planckian scattering is reached for slightly overdoped samples and does not appear to be exceeded.

Twisted bilayer graphene is a platform for strongly correlated electron dynamics where the 
carrier density and effective mass are independently tunable and measurable by quantum oscillations \cite{Cao2018a, Cao2018}. Strong $T$-linear resistivity was observed in the vicinity of the correlated insulating state at half filling \cite{PhysRevLett.124.076801, Polshyn2019}. A Drude analysis close to the `magic' twist angle, using densities and masses from quantum oscillations, found values of $1/\tau$ varying over almost an order of magnitude yet bounded by $\tau_\text{Pl}/\tau \lesssim 1.6$ \cite{PhysRevLett.124.076801}. That work furthermore noted that the observed Planckian bound in this system becomes more remarkable when contextualized by the $T$-linear resistivity of monolayer graphene, which has a scattering rate that is two orders of magnitude smaller. It has, however, been argued that strong electron-phonon scattering is expected as the magic angle is approached  \cite{PhysRevB.99.165112}, due to $E_F$ becoming small in (\ref{eq:first}).

The iron chalcogenide FeSe$_{1-x}$S$_x$ shows $T$-linear resistivity across its phase diagram above a temperature $T_1$. When superconductivity is suppressed with a magnetic field, $T_1$ collapses from around 10 K at $x=0$ to zero temperature at a critical $x_c \approx 0.16$. A Drude analysis of the $T$-linear regime, grounded in quantum oscillation data, found a Planckian lifetime $\tau \approx \tau_\text{Pl}$ across the entire range $0\leq x \leq 0.25$ \cite{Licciardello2019}.

The Drude analyses above all associate a single timescale to the metal, averaging the transport properties around the Fermi surface and on different Fermi sheets (see \S\ref{sec:drude2} for more details). It is not clear if this averaged timescale has any intrinsic meaning, see \cite{haldane} for a discussion of this point; perhaps it is best regarded as an empirical fact that a well-defined analysis procedure applied to a large range of materials consistently reveals a Planckian scattering rate. Measurements of angle-dependent magnetoresistance in principle contain information about the dependence of the transport lifetime around the Fermi surface, that can be extracted with a Boltzmann analysis of the resistivity data. This technique has been applied to cuprates with a $T$-linear scattering component in \cite{Abdel-Jawad2006,brad}. In particular, it has been reported in \cite{brad} that close to the (pseudogap) critical doping in Nd-LSCO there is an isotropic Planckian lifetime superimposed on a highly anisotropic elastic scattering rate.

An important aspect of the apparent universality that emerges from the full set of Drude analyses reviewed in this section is the range of circumstances in which $T$-linear resistivity is observed.  In tuned quantum critical systems it appears at low temperatures, extending down to below 100 mK in some cases, while in conventional metals and some cuprates it is seen up to nearly 1000 K.  This is certainly consistent with the suggestion, mentioned above,  that the linear power law and quasi-universal scattering rate may have some independence of the microscopic details of the scattering that causes it.

\subsection{Heat transport and phonons}
\label{sec:heat}

In addition to potentially scattering electrons, phonons in unconventional metals are unconventional in their own right, as was noted early on by \cite{PhysRevB.49.9073}. Furthermore, the large Lorenz ratio of cuprates at high temperatures suggests that phonons play an important role in heat transport in these materials \cite{PhysRevB.49.9073, PhysRevB.68.220503, Yan_2004, PhysRevB.72.054508, PhysRevMaterials.5.014603}. This stands in contrast to standard metals, in which heat is mostly carried by electrons, and is possible partly because in correlated systems the lower Fermi velocities reduce the electronic contribution to heat transport.
Heat transport in several cuprates has recently been revisited in the light of possible Planckian bounds \cite{Zhang5378, PhysRevB.100.241114}. The thermal diffusivity at high temperatures was found to be $D_\text{th} = \frac{1}{3} v_s^2 \tau$, with $\tau \approx 15 \tau_\text{Pl}$ and $v_s$ the speed of sound. The appearance of the sound speed, as well as the longer phonon lifetimes, is consistent with a dominant role for phonons in heat transport. It was estimated that the phonon mean free path becomes short at high temperatures --- mirroring the bad metallic behavior of the Planckian charge carriers in the same material. This led to the notion of an incoherent electron-phonon `soup' controlled by Planckian dissipation \cite{Zhang5378, PhysRevB.100.241114}.

Further supporting the role of phonons, but possibly going against the necessity of an electron-phonon soup, the behavior $D_\text{th} \sim v_s^2 \tau_\text{Pl}$ is also widely seen in crystalline insulators at high temperatures \cite{Zhang19869, Behnia_2019, Mousatov:2020tyc, PhysRevMaterials.5.014603, martelli2021thermal}. In the insulators heat is certainly carried by phonons and the $T$-linear decay rate arises from lattice anharmonicity. As with the cuprates discussed above, most crystalline insulators at high temperatures have phonons with lifetimes that are longer than $\tau_\text{Pl}$ by a factor of 10 or more. Nonetheless, the complex insulator MgSiO$_3$ \cite{PhysRevB.96.100302}, for example, seems to host genuinely Planckian phonons with short mean free paths. Thus, not only is a Planckian bound obeyed in all crystalline insulators for which the analysis has been performed, but these systems also push up against the bound.

Moving beyond insulators, heavily doped semiconductors give a conventional setting where phonons dominate heat transport in the presence of charge carriers. A careful comparative analysis of thermal diffusivity in cuprates and heavily doped semiconductors points to the importance of the electron-phonon interaction in both classes of material, and is consistent with a Planckian electron-phonon contribution to the high temperature scattering rate of electrons in cuprates \cite{Mousatov:2020auf}.

\subsection{Parallel with the conjectured viscosity bound}

The original discussion of Planckian transport in cuprates \cite{Zaanen2004} drew an implicit parallel with a lower bound on the shear viscosity over entropy density, $\eta/s \gtrsim \hbar/k_B$, conjectured in the same year \cite{Kovtun:2004de}. This connection was elaborated more explicitly in \cite{sachdevkeimer,Bruin804,Hartnoll:2014lpa, Zaanen:2015oix}, as we now explain. The most substantive evidence for the viscosity bound came, firstly, from a holographic computation in a relativistic plasma \cite{Policastro:2001yc} where the coupling strength could be taken to infinity, but the ratio $\eta/s$ remained bounded away from zero. Secondly, the quark-gluon plasma was measured to have $\eta/s \sim \hbar/k_B$ \cite{PhysRevLett.106.192301}.
In relativistic plasmas the shear viscosity over entropy density ratio controls transverse momentum diffusion according to $D_\perp = c^2 \eta/sT$.  Here $c$ is the speed of light.  Writing the diffusivity as $D_\perp = \frac{1}{3} c^2 \tau$, in this context the viscosity bound becomes the Planckian bound (\ref{eq:bound}).

While many simple non-relativistic media do obey the $\eta/s$ bound \cite{Kovtun:2004de}, the transverse momentum diffusivity in these systems is instead $D_\perp = \eta/(m n)$ with $m n$ the mass density. The non-relativistic momentum diffusivity $\eta/(m n)$ appears to be more universally bounded than $\eta/s$ in non-relativistic systems, for reasons that are unrelated to Planckian scattering \cite{Trachenkoeaba3747, Baggioli:2020lcf}. The possibility that
viscous effects are relevant for strange metal transport \cite{Zaanen:2018edk} is logically distinct from the presence of Planckian dissipation in such systems, and will need to be investigated by spatially resolved probes. For the purposes of this review we are not suggesting that viscosity and momentum diffusion per se are relevant to strange metals, but rather that Planckian dissipation may be an overarching principle leaving its fingerprint on diverse observables in diverse fields of physics.

\section{Distinction between Planckian and Mott-Ioffe-Regel bounds}
\label{sec:notMIR}

To bring out the physics of a possible Planckian bound, it is fruitful to contrast it with the widely discussed Mott-Ioffe-Regel limit \cite{RevModPhys.75.1085, hussey}. The Mott-Ioffe-Regel limit is a condition for transport in a metal to admit a description in terms of particle-like electronic excitations. The limit is most sharply formulated when the quasiparticles undergo elastic collisions, so that the single-particle states have a well-defined energy. In this case the primary requirement for a semi-classical Boltzmann description of transport is the ability to form localized wavepackets from superpositions of the single-particle Bloch states. To obtain an electronic quasiparticle with mean free path $\ell$ and Fermi wavevector $k_F$ it must be possible to form superpositions with uncertainties in position and wavevector $\Delta x \lesssim \ell$ and $\Delta k \lesssim k_F$, respectively. The uncertainty principle requires $\ell \gtrsim 1/k_F$, which is one version of the Mott-Ioffe-Regel limit. Another, weaker, version of the Mott-Ioffe-Regel limit is that the width of Bloch states making up the quasiparticle is bounded by the size of the Brillouin zone, $\Delta k \lesssim 1/a$, which leads to $\ell \gtrsim a$. Once $\Delta k \sim 1/a$ it is natural for inter-band transitions to become important (cf.~\cite{PhysRevLett.42.736}).

The Planckian bound, instead, is intrinsically tied up with inelastic scattering. This bound was also, historically, first discussed as a condition for the validity of a Boltzmann description of dynamics \cite{P2, peierls1996quantum}. Suppose that elastic collisions have been accounted for, leading to long-lived single-particle energy eigenstates given by superpositions of Bloch states. One can subsequently introduce interactions that allow energy to be transferred between these single-particle states on some inelastic timescale $\tau_\text{inel}$. This scattering leads to a width $\Delta E$ in the single-particle energy. If the scattering rate $1/\tau_\text{inel} \gtrsim k_B T/\hbar$ then the
uncertainty principle implies that $\Delta E \gtrsim k_B T$. That is, the uncertainty in the single-particle energy is greater than the width of the Fermi-Dirac distribution, inconsistent with drawing the electrons from that distribution. It follows that a Boltzmann description in terms of single-particle states requires $\tau_\text{inel} \gtrsim \tau_\text{Pl}$. This uncertainty principle perspective on the Planckian bound was more recently noted in \cite{DEVILLERS19841019} while the difference with the Mott-Ioffe-Regel limit was emphasized in \cite{Hartnoll:2014lpa}. Both limits first arise as conditions for the validity of a Boltzmann description: the Mott-Ioffe-Regel limit concerns the ability to form coherent particles from superpositions of Bloch states, with a spread in wavevector, while the Planckian limit is about the ability of quasiparticles to retain their existence in the face of inelastic many-body scattering, that causes a spread in energy.

The separation into elastic and inelastic scattering is clearest when the elastic scattering is a temperature-independent, additive contribution due to disorder: $\frac{1}{\tau} \sim \frac{1}{\tau_\text{elas}} + \frac{1}{\tau_\text{inel}}$. As mentioned in footnote \ref{foot:dis}, disorder scattering is observed to give such an additive contribution to the resistivity in $T$-linear transport regimes of weakly disordered cuprates.
This (presumably) elastic contribution, responsible for the residual resistivity at $T=0$, can then be subtracted off, as was done in the Drude analyses discussed in \S\ref{sec:drude}. This subtraction is important because the elastic lifetime is not subject to a Planckian bound. As noted in \S\ref{sec:intro2} and in the preceding paragraphs, single-particle Bloch states are not energy eigenstates once lattice translation invariance is broken, and can correspondingly be short-lived even in non-interacting systems. The residual resistivity then arises because simple transport observables interrogate the system in the `wrong' (Bloch, non-eigenstate) single-particle basis. In general, however, it is not possible to perform such a subtraction in a model-independent way. For example, in \S\ref{sec:elas} we recall that scattering from phonons at high temperatures is elastic with a temperature-dependent scattering rate. The difficulty with isolating the `inelastic' part of scattering motivates a more many-body perspective on this problem that we will turn to shortly.

It is well-known that the Mott-Ioffe-Regel limit can be violated in non-saturating `bad' metals \cite{PhysRevLett.74.3253, RevModPhys.75.1085, hussey}. This has led to an extensive effort to characterize transport beyond the Boltzmann paradigm, without a notion of a mean free path. For example, \cite{PhysRevB.73.035113} developed a `statistical' description of non-quasiparticle high temperature transport while \cite{PhysRevLett.122.216601} gives a non-quasiparticle perspective on the distinction between elastic and inelastic dynamics.
The proposal that we set out to examine in this review is that some version of a Planckian bound holds independently of the existence of well-defined quasiparticles. An important piece of this proposal is that the inelastic lifetime of quasiparticles is an avatar of a more basic many-body timescale: the equilibration time.

\subsection{Equilibration time}
\label{sec:eq}

In thermal equilibrium much of the complexity of many-body dynamics is universally subsumed into a small number of variables such as the temperature $T$ and charge density $n$. The next most universal question one can ask concerns the approach in time to thermal equilibrium. Given a perturbation of the thermal state, how quickly can equilibrium be re-established? More precisely, the interesting question to ask is how quickly {\it local} thermal equilibrium can be re-established. The answer to this latter question will define an intrinsic timescale that does not depend on the size of the system. In this section we aim to give a brief self-contained introduction to the key concepts of
equilibration length and local equilibration time. These can be thought of as generalizations of the notion of inelastic quasiparticle mean free path and lifetime, respectively. In the following section we explain that the equilibration length and time are bounded by basic properties of the thermal state. That is to say, they are subject to constraints that transcend any given description of transport, quasiparticles or no quasiparticles. It is possible, therefore, that at least some of the Planckian timescales appearing in electronic transport reflect a potential Planckian bound on the local equilibration time.

If a many-body system has microscopic interactions that are local in space, then its thermal state has an `equilibration length' $\ell_\text{eq}$. Physically, $\ell_\text{eq}$ is the linear size of the smallest region that is able to maintain thermal equilibrium with itself. For example, in a quasiparticle system $\ell_\text{eq}$ will be a few times the (inelastic) mean free path. Different regions of size $\ell_\text{eq}$ can independently establish different local temperatures $T(x)$ and densities $n(x)$. The time taken for such a region to reach local thermal equilibrium will define the equilibration time $\tau_\text{eq}$.

After $\tau_\text{eq}$ the locally thermalized $T(x)$ and $n(x)$, which vary on scales greater than $\ell_\text{eq}$, relax to global thermal equilibrium via diffusion of heat and charge.\footnote{This is the case insofar as the electronic degrees of freedom form a closed quantum system. Once the lattice degrees of freedom are included, there will be linearly dispersing sound modes in addition to charge diffusion. These will dominate global equilibration at sufficiently long distances. The role of the lattice in global equilibration raises the question of whether it is consistent to restrict attention to the electronic subsystem in considering local equilibration. To address this question we can ask about the local equilibration of the combined electron-lattice system. In a conventional Bloch-Gr\"uneisen description, above the Debye temperature at least, the exchange of energy between the electronic and lattice subsystems is the bottleneck for local equilibration of the combined system \cite{PhysRevLett.59.1460}. Consistently with this picture, time-resolved photoemission experiments in cuprates have suggested that the electronic subsystem locally equilibrates on a faster timescale than energy is exchanged with the lattice \cite{PhysRevLett.99.197001, DalConte2015, Rameau2016, Konstantinovaeaap7427}. This fact, in turn, suggests that it is indeed consistent to consider the equilibration time of
the electronic subsystem, viewed as a closed quantum system. However, even if interactions with the lattice do not transfer energy out of the electronic subsystem efficiently
they can still, as discussed in \S\ref{sec:elph}, produce important collision processes that strongly degrade electronic single-particle Bloch states. When this scattering is significant (as in conventional metals), then electronic equilibration must be understood to occur in the presence of these external processes, which can be viewed as loosely analogous to disorder scattering. \label{foot:equil}} We discuss diffusion, including possible bounds on the diffusivity, in Appendix \ref{sec:diff} below. Global thermal equilibrium is not reached until diffusion extends across the whole system at the (very long) Thouless time $\tau_\text{Th} = L^2/D$, with $L$ the spatial extent of the system and $D$ the diffusivity. In contrast to the local equilibration time, this is not an intrinsic timescale. The overall picture is that there are fast, exponential in time, processes leading to local thermal equilibrium followed by slower power law processes describing the diffusive approach to global equilibrium.\footnote{However, fluctuations in the late time diffusive dynamics will mix with the decay of generic local quantities, leading to power law rather than exponential decay \cite{Chen-Lin:2018kfl, Delacretaz:2020nit}. Disentangling the fluctuation effects remains an important open problem.} The local equilibration time $\tau_\text{eq}$ characterizes the slowest of the fast processes, the last decay to occur before the dynamics becomes diffusive.

\subsection{Constraints on equilibration as possible routes to a Planckian bound}

The notion that $\tau_\text{eq}$ may be subject to a Planckian bound is supported by the fact that local equilibration cannot happen arbitrarily quickly, as interactions are needed to distribute energy between the various degrees of freedom. A simple many-body uncertainty principle argument, given in Appendix \ref{sec:uncertain}, constrains how fast a region of size $\ell_\text{eq}$ can evolve in terms of the largest local coupling $J$ in a lattice Hamiltonian with locally bounded energies:
\be\label{eq:tJ}
\tau_\text{eq} \gtrsim \frac{\hbar}{J} \,.
\ee
This is a rather microscopic bound, that is weaker than the Planckian bound at temperatures well below the `bandwidth' $J$. The main point, however, is that the equilibration rate is fundamentally bounded in this context. Establishing a Planckian bound would therefore be a question of improving this constraint, rather than starting from scratch. Furthermore, if $T$-linear scattering survives up to temperatures that are some fraction of the bandwith, as is seen in high temperature quantum Monte Carlo studies of the Hubbard model \cite{Huang987},  then (\ref{eq:tJ}) will be violated once $k_BT \sim J$ unless there is a Planckian bound on the numerical prefactor of the $T$-linear scattering. Thus, (\ref{eq:tJ}) may be a hint of a Planckian bound. Indeed, it is likely that it will be possible to strengthen the many-body uncertainty principle argument at low temperatures and to Hamiltonians that are not locally bounded. For example, the mathematical origin of the Planckian bound on the Lyapunov exponent --- see \S\ref{sec:lyapunov} below --- has the flavor of a many-body uncertainty principle. In essence the proof works by showing that functions with an `energy width' (formally, a periodicity in imaginary time) set by the temperature cannot vary on faster than Planckian timescales \cite{Maldacena:2015waa}. Further discussion of the role of the Planckian time in thermal equilibration can be found in \cite{Goldstein_2015, NUSSINOV2020114948, Nussinov:2021fgc, Pappalardi:2021ahe}.

A distinct fundamental constraint on local equilibration was articulated in \cite{Delacretaz:2018cfk,dela}. This constraint relates more naturally to $\ell_\text{eq}$ rather than $\tau_\text{eq}$. We will correspondingly see that it leads to a version of the Mott-Ioffe-Regel limit that is formulated without reference to quasiparticles, potentially providing insight into the nature of bad metals. The starting point is that locally thermalized regions of finite size $\ell_\text{eq}$ will experience thermal fluctuations. The smaller the region, the larger the fluctuations. Intuitively, these fluctuations cannot become too big if local equilibrium is to be meaningful and this leads to a lower bound on $\ell_\text{eq}$. Simple estimates given in Appendix \ref{sec:fluc} suggest that for fermions with density $n$ in $d$ dimensions one must have
\be\label{eq:nell}
n \ell_\text{eq}^d \gtrsim 1\,.
\ee
This is the intuitive statement that for a region to be able to self-thermalize it should contain more than one particle.\footnote{For conformal field theories, arising at bosonic quantum critical points, the corresponding constraint is $s \ell_\text{eq}^d \gtrsim 1$, with $s$ the entropy density \cite{Delacretaz:2018cfk}. In a conformal field theory $s \sim (k_B T/\hbar c)^d$ and $\ell_\text{eq} \sim c \tau_\text{eq}$, with $c$ the effective speed of light. The fluctuation bound therefore becomes the Planckian bound $\tau_\text{eq} \gtrsim \hbar/(k_B T)$ in this particular case, consistent with the explicit computations in \cite{subirbook}.\label{foot:cft}}

For degenerate fermions $n \sim k_F^d$ and hence (\ref{eq:nell}) becomes a version of the Mott-Ioffe-Regel limit, $k_F \ell_\text{eq} \gtrsim 1$. In bad metals
some lengthscale $\ell$ extracted from transport strongly violates the Mott-Ioffe-Regel limit, while $\ell_\text{eq}$ must still obey (\ref{eq:nell}). It follows that in these cases $\ell \ll \ell_\text{eq}$. This hierarchy suggests that the transport length $\ell$ is not directly related to equilibrating electronic dynamics in bad metals. On this point it is interesting to note that measurements of spin \cite{Sommer2011, PhysRevLett.118.130405, threview} and momentum \cite{Cao58, PhysRevLett.124.240403} transport in degenerate Fermi liquids tuned to unitarity using trapped ultracold atoms {\it do} show saturation at the Mott-Ioffe-Regel limit. However, once the ultracold atoms are placed in a rigid optical lattice then non-saturating behavior can arise \cite{Brown379, Xu2019}. This possibly suggests that the finite bandwidth due to the lattice is an essential ingredient to produce bad metals. In condensed matter systems, scattering by lattice vibrations at high temperatures furthermore gives a natural mechanism to disassociate equilibration and transport lengthscales, as discussed in footnote \ref{foot:equil} and also in \S\ref{sec:elas} below.

\section{Quasiparticle lifetime in a metal}
\label{sec:single}

While the equilibration time is a conceptually useful quantity to have in mind, it has not been directly measured in condensed matter systems. However, many existing measurements do reveal characteristic timescales in a metal. In this section we will discuss single-particle electronic dynamics, as probed by e.g.~angle-resolved photoemission \cite{RevModPhys.75.473, sunko2019angle}, while the following \S\ref{sec:transportime} will consider transport dynamics. As we described in \S\ref{sec:notMIR}, elastic scattering produces a broadening $\Delta k$ in wavevector, due to momentum-position uncertainty. An intrinsic broadening $\Delta \omega$ in frequency requires, in contrast, a single-particle energy-time uncertainty from inelastic many-body equilibration. However, photoemission (and transport) probe Bloch states with definite wavevectors. Due to the elastic broadening $\Delta k$, Bloch states are a superposition of single-particle eigenstates. For this reason, these measurements will observe a broadening $\Delta \omega$ in frequency that is unrelated to many-body equilibration. Therefore, in order for the measured electronic timescales to give reasonable estimates of the equilibration time, elastic scattering must be negligible or it must be possible to subtract it out. Furthermore, the electrons must be able to self-equilibrate prior to equilibrating with a larger lattice system (see footnote \ref{foot:equil}).
Once these conditions are met,
the unstable state created by exciting a single electron or an electron-hole pair will decay by all channels that are available, including (unless excluded by some symmetry) the slowest of the fast processes that determines the local equilibration time. This fact provides a link between the Planckian observations summarized in \S\ref{sec:intro} and potential bounds on dissipation discussed in \S\ref{sec:notMIR}.

As we have discussed in \S\ref{sec:notMIR}, single-particle states with a Planckian lifetime are on the verge of dissolving into an inherently many-body state that is beyond a Boltzmann-type description of dynamics. The idea of a Planckian quasiparticle lifetime may therefore seem to be inconsistent.  There are two points to make here. The first is that a timescale can still be usefully estimated from the width of peaks in single-particle spectral functions, independently of whether or not these peaks are the correct starting point to describe collective processes such as transport. Secondly, there appear to be circumstances where a quasiparticle description survives in $T$-linear transport regimes. This can occur due to simplifications in the dynamics (such as the hierarchy between phonon and electronic energy scales, discussed in \S\ref{sec:phdeb}) or because a temperature-dependent effective mass leads to sub-Planckian decay rates even while the resistivity is $T$-linear (as is the case, marginally, in the marginal Fermi liquid discussed in \S\ref{sec:MFL}). In this section, therefore, we will characterize quasiparticle timescales starting from the single-particle electronic Green's function. Concise introductions to the aspects of Green's functions that we will use may be found in e.g. \cite{RevModPhys.75.473, sunko2019angle}.

The full electronic retarded Green's function
is
\be\label{eq:Green}
G^R(\omega,k) = \frac{1}{\omega - \epsilon_k - \Sigma(\omega,k)} \,.
\ee
Here $\epsilon_k$ is the bare electronic dispersion and $\Sigma(\omega,k)$ is the self-energy. There are no assumptions going into (\ref{eq:Green}), although writing the Green's function in this form is only likely to be useful if the self-energy has a simple structure such as in the case of quasiparticles discussed shortly. As a counterexample,
in a Luttinger liquid the electronic Green’s function contains dispersing power law singularites $G^R(\omega,k) \sim (\omega - v k)^\delta$ \cite{Schonhammer2004}, which are not sensibly put in the form (\ref{eq:Green}).
The full Green's function is, in principle, directly accessible through angle-resolved photoemission experiments \cite{RevModPhys.75.473, sunko2019angle}. The self-energy can be split into real and imaginary parts $\Sigma = \Sigma' + i \Sigma''$ that obey the Kramers-Kronig relation
\be\label{eq:KK}
\Sigma'(\omega,k) = \frac{ {\mathcal P}}{\pi} \int_{-\infty}^\infty \frac{\Sigma''(\omega',k)}{\omega' - \omega} d\omega' \,.
\ee
Here ${\mathcal P}$ denotes the principal value of the integral.

At low energies, compared to the underlying bare electronic energy scales, often only the spectral weight close to the Fermi surface is important. When this is the case, one can expand the dispersion and self-energy about the Fermi surface to obtain a simpler `quasiparticle' Green's function with a Lorentzian form, writing
\be\label{eq:expand}
\epsilon_k = v_F \cdot k_\perp + \cdots \,, \quad
\Sigma(\omega,k) = i \Sigma''(0,k_F) + \w \pa_\omega \Sigma'(0,k_F) + k_\perp \cdot \nabla_k \Sigma'(0,k_F) + \cdots \,.
\ee
Here $k_\perp$ is a vector orthogonal to the Fermi surface with magnitude the distance to the Fermi surface, $v_F$ is the Fermi velocity and $k_F$ the Fermi wavevector. All of these quantities, in general, can vary around the Fermi surface. Let us denote $\Sigma_o = \Sigma(0,k_F)$ in the following. Using the expansion (\ref{eq:expand}) in the Green's function (\ref{eq:Green}) gives
\be\label{eq:Green2}
G^R(\omega,k) = \frac{Z}{\omega - v_F^\star \cdot k_\perp + i/\tau} \,,
\ee
where
\be\label{eq:tau}
Z = \frac{1}{1 - \pa_\omega \Sigma_o'} \,, \qquad v_F^\star = Z \left(v_F + \nabla_k \Sigma_o'\right) \,, \qquad  \frac{1}{\tau} = - Z \Sigma_o'' \,.
\ee
These quantities will also, in general, vary around the Fermi surface. The assumptions leading to (\ref{eq:Green2}) are that low energy electronic physics occurs close to a Fermi surface and that all quantities admit an analytic expansion in small $\omega$ and $k_\perp$ close to the Fermi surface. The latter assumption will be true at nonzero temperatures. In e.g. a quantum critical regime, however, non-analytic behavior can develop at relatively low frequencies $\hbar \omega \sim k_B T$. This can have important consequences: if the self-energy has a nontrivial frequency dependence over the width $1/\tau$ of the single-particle peak --- i.e. the peak is not a Lorentzian --- then it is not possible to deduce an unambiguous unique lifetime for the quasiparticle.

In (\ref{eq:Green2}) the immediately physical timescale is $\tau$. Upon Fourier transforming the Green's function this is seen to be the time over which a quasiparticle excitation decays. We see in (\ref{eq:tau}) that in general this time is not equal to the imaginary part of the self-energy at the Fermi surface $\Sigma_o''$, due to the factor of the quasiparticle residue $Z$.

The Kramers-Kronig relation (\ref{eq:KK}) implies that (being careful with taking the derivative of the principal value ${\mathcal P}$)
\be\label{eq:overZ}
\frac{1}{Z} = 1 - \left. \frac{\pa \Sigma'}{\pa \omega} \right|_{\omega = 0} = 1 -  \frac{{\mathcal P}}{\pi} \int_{-\infty}^\infty \frac{\Sigma''(\omega,k_F) - \Sigma_o''}{\omega^2} d\omega \,.
\ee
Note that $\Sigma'' < 0$ in the conventions being used. The physical content of (\ref{eq:overZ}) is that even while a low frequency electron cannot be efficiently scattered by high frequency degrees of freedom, these `fast' modes do renormalize the effective medium through which the electron moves. Both $\Sigma_o''$ and $Z$ appear in the physical single particle decay rate $1/\tau$ in (\ref{eq:tau}).

In several important circumstances the modes scattering the electrons have a slower characteristic velocity than the electrons themselves, resulting in
$\nabla_k \Sigma'_o \ll v_F$ in (\ref{eq:tau}). For example, this is the case for scattering by acoustic phonons because the sound speed is less than the Fermi velocity. In such cases the shift by $\nabla_k \Sigma'_o$ may be dropped and the Fermi velocity is simply multiplicatively renormalized by $Z$. We may then also identify this factor as a mass renormalization
\be
\frac{m_\star}{m} = \frac{|v_F|}{|v_F^\star|} = \frac{1}{Z} \,.  \qquad \qquad (\nabla_k \Sigma'_o \ll v_F)
\ee
The expression for $1/\tau$ in (\ref{eq:tau}) does not depend upon this additional simplification.

It is instructive to see how the physical timescale $\tau$ relates to the Planckian bound in two simple and widely discussed models of $T$-linear resistivity: electron-phonon systems and the marginal Fermi liquid. 
We do this in the following few sections. In both cases we will see that at the lower end of the temperature range showing $T$-linear resistivity the physical lifetime automatically obeys the Planckian bound, while at high temperatures a Planckian bound on the lifetime requires a bound on the magnitude of a dimensionless coupling constant that controls the strength of microscopic scattering.

\subsection{Electron-phonon interactions revisited}
\label{sec:ep2}

In this section we continue the discussion of electron-phonon interactions from \S \ref{sec:elph}, now accounting for the renormalization of the electronic quasiparticle residue due to the phonons. We have just seen in (\ref{eq:tau}) that consideration of this effect is important to obtain the physical lifetime of the electrons, and will modify the previous expression (\ref{eq:first}) for $1/\tau$.

To illustrate the physical points in a simple and explicit setting, consider scattering of electrons by an Einstein phonon at frequency $\omega_\text{ph}$.\footnote{The physics is qualitatively similar for dispersive phonons. The main difference is that $T$-linear scattering occurs above roughly $\frac{1}{3} T_\text{BG}$, where the Bloch-Gr\"uneisen temperature for acoustic phonons with sound speed $v_s$ is $k_B T_\text{BG} \sim \hbar v_s k_F$. This can be lower than the Debye temperature if the Fermi surface is small, such that $k_F \ll 1/a$. See e.g. \cite{PhysRevB.99.085105} for a recent discussion.} The imaginary part of the self-energy is (from e.g. \S 7.4.2 of \cite{mahan})
\be\label{eq:Sphonon}
\Sigma''(\omega) = \pi \l_\text{e-ph} \, \omega_\text{ph} \sum_\pm \frac{\pm f(\omega \mp \omega_\text{ph}) b(\pm \omega_\text{ph})}{f(\omega)} \,.
\ee
Here $\l_\text{e-ph}$ is a dimensionless effective electron-phonon coupling, incorporating a factor of the density of states at the Fermi surface --- this is essentially the prefactor in (\ref{eq:first}). The factor of $\pi$ in (\ref{eq:Sphonon}) gives the conventional normalization of $\l_\text{e-ph}$. The functions $f(x) = 1/(e^{\hbar x/k_B T} + 1)$ and $b(x) = 1/(e^{\hbar x/k_B T} - 1)$ are the Fermi-Dirac and Bose-Einstein distributions, respectively. The frequencies $\omega$ and $\omega_\text{ph}$ are assumed to be small compared to the Fermi energy, and there is no momentum dependence in the self-energy.

\begin{figure}[h]
    \centering
    \includegraphics[width=0.65\textwidth]{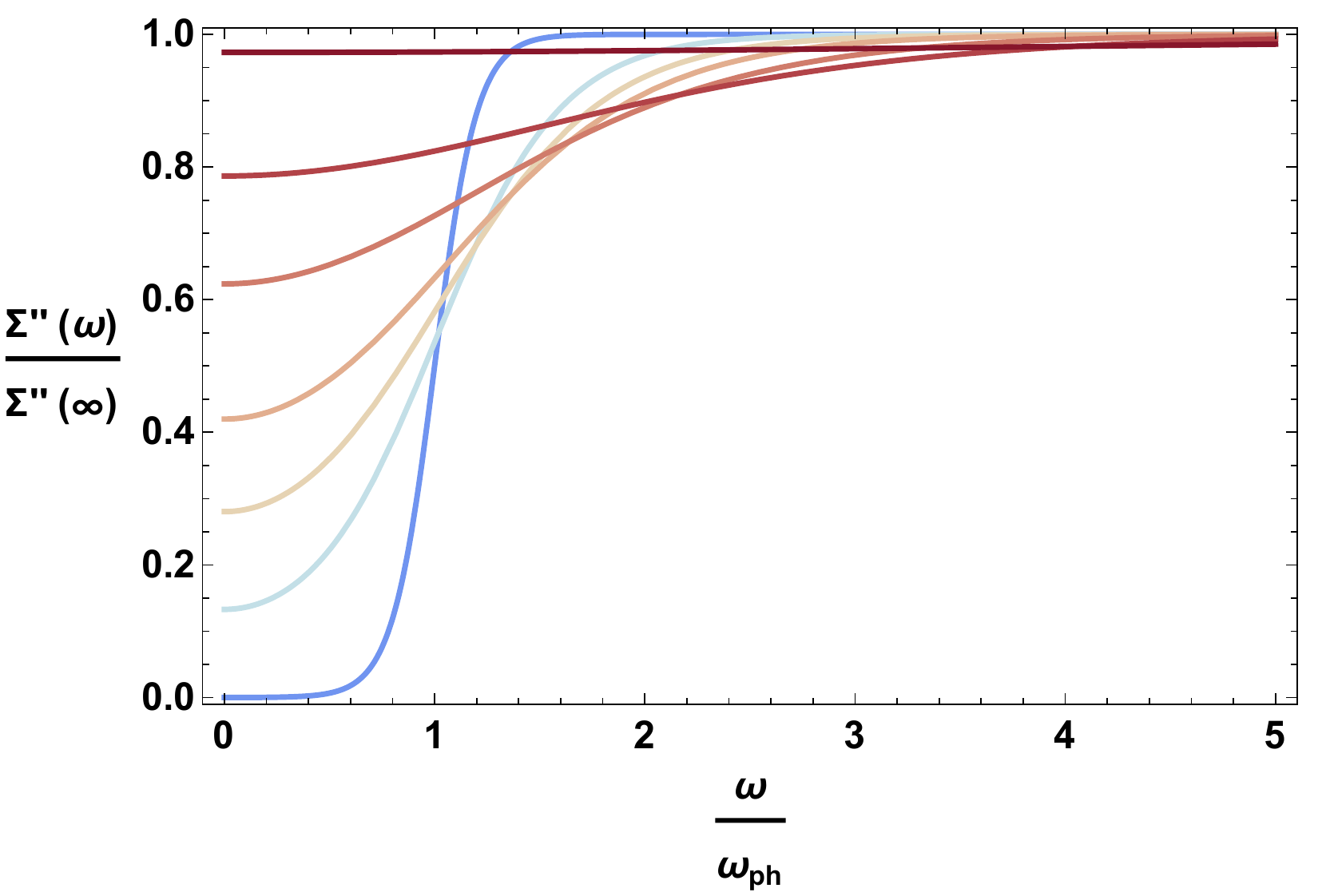}
    \caption{The imaginary part of the self-energy (\ref{eq:Sphonon}) due to scattering by an Einstein phonon, normalized at large frequencies. From bottom to top, temperatures are $T/T_\text{ph} = 0.1,0.3,0.4,0.5,0.7,1,3$. At low temperatures no scattering occurs below the frequency $\omega_\text{ph}$, while at high temperatures all frequencies are scattered.}
    \label{fig:sigmaph}
\end{figure}

At low temperatures $T \ll T_\text{ph}$ (with $k_B T_\text{ph} = \hbar \omega_\text{ph}$) the function $\Sigma''(\omega)$ in (\ref{eq:Sphonon}) vanishes for $|\omega| < \omega_\text{ph}$ and is constant outside this range. As temperature is increased the dip at small $\omega$ becomes less pronounced until, at $T \gg T_\text{ph}$, the function is just flat (see Fig.~\ref{fig:sigmaph}).
This is physically transparent: at low temperatures the electron needs to have frequency $\omega_\text{ph}$ to be able 
to emit a phonon. At high temperatures all electrons are able to do this. The unrenormalized scattering rate at the Fermi surface following from (\ref{eq:Sphonon}) is
\be\label{eq:csch}
- \Sigma_o'' = 2 \pi \l_\text{e-ph} \, \omega_\text{ph} \, \text{csch} \frac{T_\text{ph}}{T} \,,
\ee
which is $T$-linear for $T \gtrsim \frac{1}{3} T_\text{ph} \equiv T_o$, leading to a $T$-linear resistivity shown in Fig.~\ref{fig:phononrho} below.

To obtain the physical scattering rate $1/\tau$ we must multiply (\ref{eq:csch}) by the residue $Z$, which can be computed from $\Sigma''(\omega)$ using the Kramers-Kronig integral (\ref{eq:overZ}). The range of the integral can be taken to infinity when $\omega_\text{ph}$ and $T$ are well below the Fermi energy, because the integral converges at high frequency. We will discuss separately temperatures close to the lower end of the $T$-linear regime, $T \sim T_o$, and high temperatures $T \gg T_o$.

\subsection{Planckian bound around the phonon energy scale}
\label{sec:phdeb}

At the lower end of the $T$-linear temperature range, $T=T_o$, the Kramers-Kronig integral gives the physical inverse lifetime
\be\label{eq:lowTph}
\frac{1}{\tau(T_o)} \, \approx \, \frac{0.63\,  \l_\text{e-ph} \omega_\text{ph}}{1 + 2.3 \, \l_\text{e-ph}} \, \leq \, 0.81 \times \frac{k_B T_o}{\hbar} \sim \frac{1}{\tau_\text{Pl}(T_o)} \,.
\ee
The numerical factors here are obtained using the illustrative Einstein phonon expression (\ref{eq:Sphonon}) and should not be quantitatively compared against experiment. The important point is that the physical lifetime of electrons scattering from phonons is naturally Planckian bounded at the lower end of the temperature range showing $T$-linear scattering (and indeed for all $T \lesssim T_\text{ph}$). The expression (\ref{eq:lowTph}) shows that, at these temperatures, the possible hierarchies of energy scales discussed in \S\ref{sec:elph} which would lead to a large $\lambda_\text{e-ph}$ do not, in fact, lead to super-Planckian decay rates. Many conventional metals have $2 \pi \lambda_\text{e-ph}$ of order one and therefore do not probe this effect. However, as noted in \S\ref{sec:elph}, conventional superconductors such as Pb and Nb have a bare $2 \pi \lambda_\text{e-ph}$ that is an order of magnitude larger \cite{allen}. As is suggested by (\ref{eq:lowTph}), significant low temperature mass renormalization in these cases ensures that the physical lifetime at $T \lesssim T_\text{ph}$ is nonetheless within an order one numerical factor of the Planckian time \cite{Bruin804}.

In \S\ref{sec:notMIR} we explained that Planckian single-particle lifetimes suggest a breakdown of the Boltzmann description of dynamics, raising the question of whether
the above simple scattering estimate of a Planckian lifetime is self-consistent. Furthermore, if $\lambda_\text{e-ph}$ becomes too large then higher order scattering processes may be important.
It has been argued by Prange and Kadanoff that the standard Boltzmann transport equations apply to electronic excitations that have been strongly broadened by phonon scattering \cite{PhysRev.134.A566}. Their arguments are based on the hierarchy of energy scales $T_\text{ph} \ll T_F$, which means that while characteristic phonon wavevectors are comparable to electronic ones, phonon energies are much smaller. This suggests several simplifying approximations, first noted by Migdal \cite{migdal1958interaction}, including the fact that 
the electron self-energy depends more strongly on energy (where it varies on the scale $T_\text{ph}$) than momentum (where it varies over lattice scales). The Migdal approximation has, however, recently been shown to give qualitatively incorrect results already at relatively small values of the electron-phonon coupling in the context of Migdal-Eliashberg theory \cite{PhysRevB.97.140501, CHUBUKOV2020168190}. The missing physical ingredient is the tendency towards polaron formation at large coupling. Electron-phonon transport theory in this regime ($T \sim T_\text{ph}$ with $2 \pi \lambda_\text{e-ph} \sim 1$) should be revisited in the light of those results. Recent works have studied Planckian transport by polarons \cite{PhysRevLett.123.076601, PhysRevResearch.1.033138}.

\subsection{Emergent elastic scattering at high temperatures}
\label{sec:elas}

At high temperatures $T \gg T_\text{ph}$, there are two important differences compared to the discussion above. Firstly, phonons do not efficiently renormalize the quasiparticle residue. This is because the phonon frequency, bounded above by $\omega_\text{ph}$, is not fast enough to renormalize the medium experienced by the hot electrons at these temperatures. Mathematically, from the Kramers-Kronig relation, it is found that at temperatures $T \gg T_\text{ph}$ the correction to $1/Z$ due to phonons vanishes like $1/T^2$. The physical decay rate is then simply the bare one
\be\label{eq:lam}
\left. \frac{1}{\tau} \right|_{T \gg T_\text{ph}} \approx - \Sigma_o'' \, \approx \, 2 \pi \l_\text{e-ph}  \times \frac{k_B T}{\hbar} \,.
\ee
A Planckian bound on the physical lifetime at these higher temperatures would, then, require the coupling itself to be bounded: $2 \pi \l_\text{e-ph} \lesssim 1$. We have already noted that this bound can be somewhat violated, even while it is saturated by many conventional metals. However, and this is the second point, the decay of single-particle Bloch states due to phonon scattering is not in fact subject to a Planckian bound in this high temperature regime, as we now explain. It follows that $\lambda_\text{e-ph}$ need not be bounded, at least not for this reason.\footnote{However, the electron-phonon coupling controls the equilibration rate of the combined electronic and lattice system when the bottleneck for equilibration is the exchange of energy between the two subsystems \cite{PhysRevLett.59.1460}. In these circumstances a Planckian bound on the full equilibration time will upper bound $\lambda_\text{e-ph}$, as is noted e.g.~on page 54 of Abrikosov's textbook \cite{abrikosov}. Furthermore, it has recently been shown that bipolaron formation prevents metallic transport with a large electron-phonon coupling \cite{murthy2021stability}. Finally, it may also be that large values of the electron-phonon coupling are excluded for other reasons, such as `quantum engineering' constraints on the stability of the crystal that are extrinsic to the electron-phonon dynamics. \label{foot:pol}} The argument we give essentially follows Peierls (see e.g. \cite{P1} or \S6.8 of \cite{peierls1996quantum}), who attributes it to Landau.

At sufficiently high temperatures scattering by phonons is approximately elastic in the sense that the highly energetic electrons do not lose a significant fraction of their energy in emitting or absorbing a phonon with energy $\hbar \omega_\text{ph}$. Thus the phonons effectively create a static disordered background potential for the electrons. Electronic motion can be described in terms of the single-particle eigenstates for motion in this effective potential, which are superpositions of Bloch states. This superposition fully accounts for the elastic part of the scattering. The remaining inelastic part will then cause interactions between these single-particle states. However, when the scattering is predominantly elastic, the energy spread in the single-particle states induced by the inelastic many-body interactions will be small and in particular less than the width of the Fermi-Dirac distribution. As for the elastic disorder scattering discussed in \S\ref{sec:notMIR}, then, the only constraint for a Boltzmann description is the ability to form coherent wavepackets. There is no requirement of a Planckian bound and, indeed, a controlled large-$N$ model of electron-phonon interactions with transport lifetimes violating the Planckian bound in such a regime was constructed in \cite{PhysRevB.93.075109}.

\begin{figure}[h]
    \centering
    \includegraphics[width=0.62\textwidth]{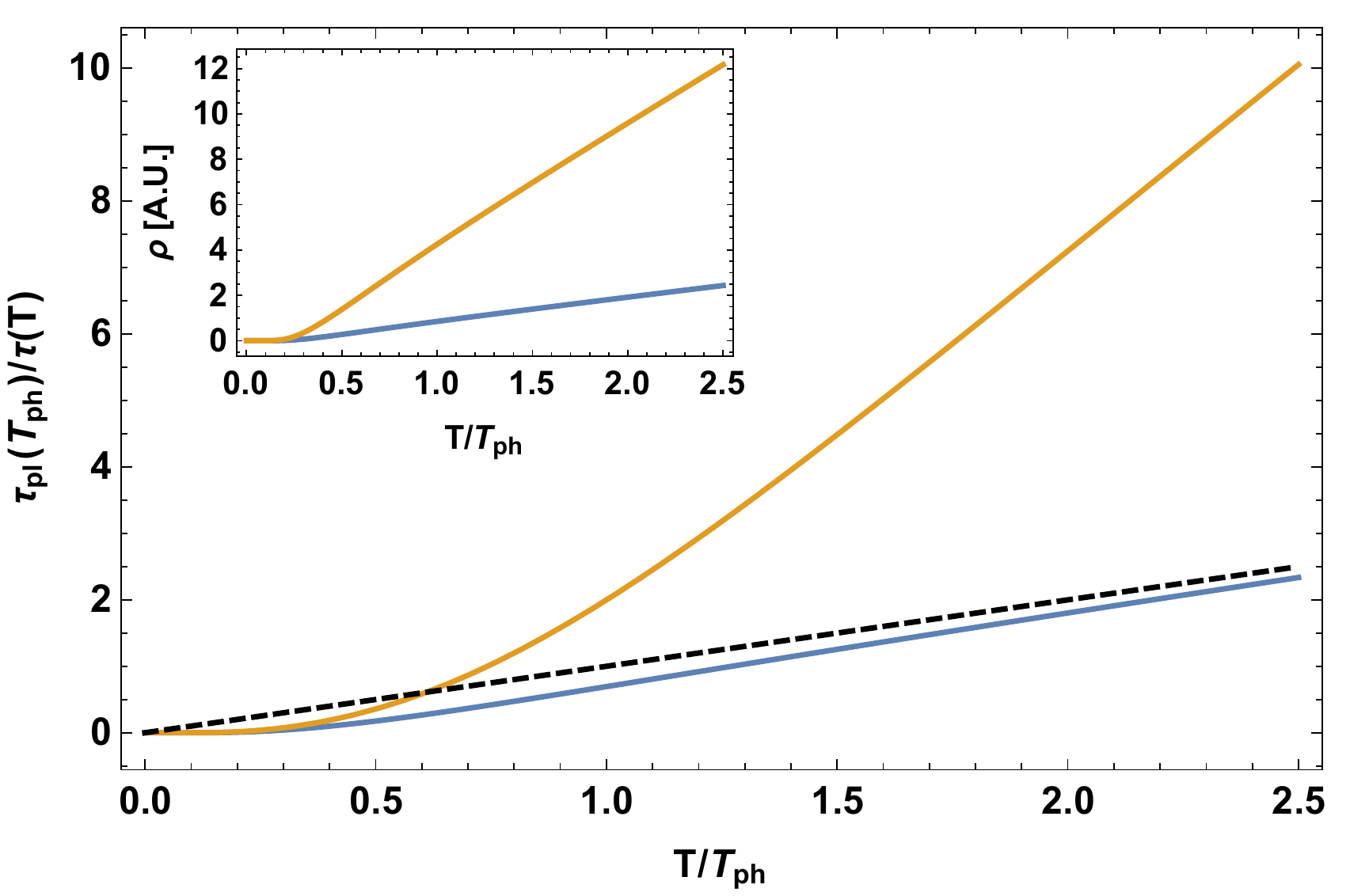}
    \caption{The inverse quasiparticle lifetime as a function of temperature due to scattering by an Einstein phonon, following from (\ref{eq:Sphonon}). The electron-phonon coupling is $2 \pi \lambda_\text{e-ph} = 5$ (top curve) and $2 \pi \lambda_\text{e-ph} = 1$ (bottom curve). The dashed line shows the Planckian rate. The inset shows the corresponding resistivity. The resistivity is $T$-linear for $T \gtrsim \frac{1}{3} T_\text{ph}$ with a slope proportional to the coupling. Even with the larger coupling, however, the inverse lifetime becomes sub-Planckian for $T \lesssim T_\text{ph}$ due to a temperature-dependent mass renormalization.}
    \label{fig:phononrho}
\end{figure}

Fig.~\ref{fig:phononrho} illustrates both the intermediate and high temperature regimes discussed.
The resistivity is $T$-linear for $T \gtrsim \frac{1}{3} T_\text{ph}$ and is insensitive to the increasingly
elastic nature of the scattering (cf.~Fig.~\ref{fig:martin}).
This is because, as we recall in \S\ref{sec:transportime} below, the renormalization of the effective mass, and hence the factor of $(1 + 2.3 \, \l_\text{e-ph})^{-1}$ in (\ref{eq:lowTph}), cancels from the resistivity.
The physical quasiparticle lifetime is, however, sensitive to mass renormalization. When the electron-phonon coupling is large, the
high temperature elastic scattering (\ref{eq:lam}) can be super-Planckian, but
mass renormalization drives the scattering rate towards or below the Planckian bound in the inelastic regime $T \lesssim T_\text{ph}$, as we saw in (\ref{eq:lowTph}).

\subsection{Quasiparticle lifetime in a marginal Fermi liquid}
\label{sec:MFL}

The marginal Fermi liquid (MFL) was originally postulated as a phenomenological description of the cuprates \cite{PhysRevLett.63.1996}.  However, it has now been shown to result from a number of explicit microscopic models and, as described below, its predicted relationship between $T$-linear resistivity and logarithmic electronic heat capacity has been widely observed.  It has self-energy
\be\label{eq:MFL}
\Sigma''(\omega) = - \lambda \max(|\omega|,k_B T/\hbar) \,.
\ee
Here $\lambda$ is again a dimensionless coupling constant. The self-energy has no strong dependence on momentum, a feature that the MFL shares with the case of scattering by phonons that we have just discussed. We recalled in \S\ref{sec:qc} that a MFL can be obtained by scattering fermions from a critical bosonic mode with low energy spectral weight spread over a range of momenta, avoiding small-angle suppression of the scattering in the transport lifetime.

\begin{figure}[h]
    \centering
    \includegraphics[width=\textwidth]{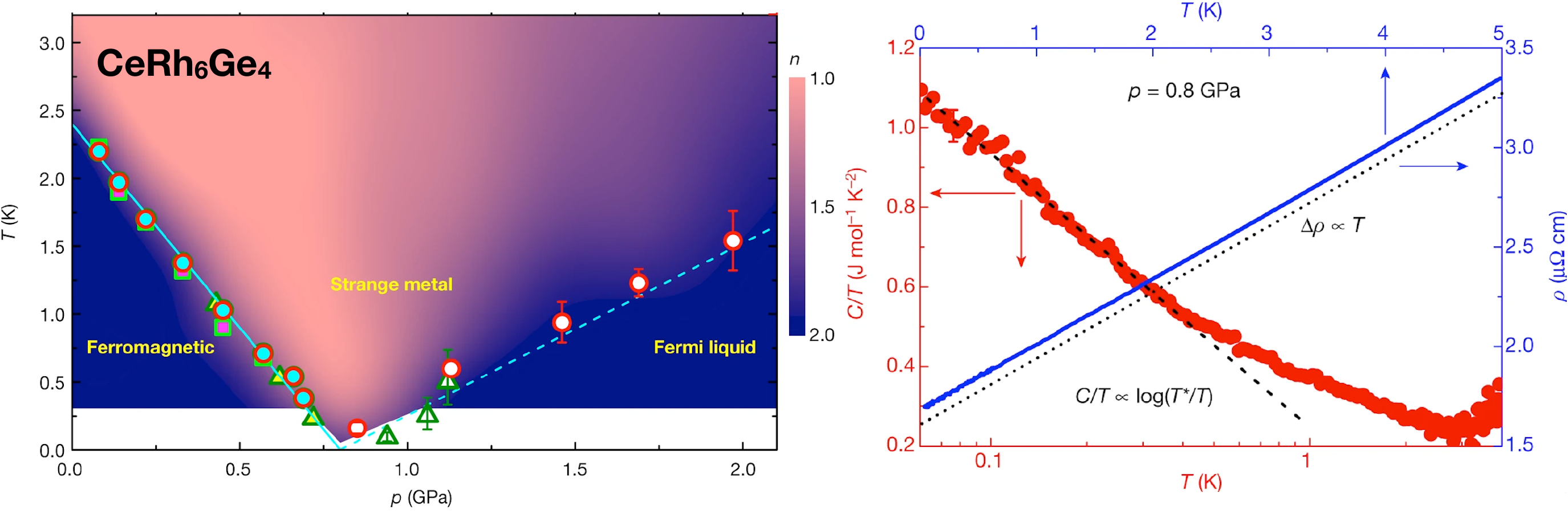}
    \caption{$T$-linear resistivity and logarithmically enhanced specific heat upon tuning the ferromagnetic heavy fermion CeRh$_6$Ge$_4$ to a quantum critical point \cite{Shen2020}. In the left hand plot, $n$ is the logarithmic derivative of the resistivity (i.e. $n=1$ is $T$-linear) as a function of pressure and temperature. The right hand plot shows the resistivity and specific heat as a function of temperature at the critical pressure.}
    \label{fig:MFL}
\end{figure}

The form (\ref{eq:MFL}) is assumed to hold over the range $0 \leq |\omega| < \omega_\star$, for some cutoff frequency $\omega_\star$. The corresponding cutoff temperature is $T_\star = \hbar \omega_\star/k_B$, with $T$-linear resistivity found {\it below} the temperature $T_\star$. As
we emphasized in \S\ref{sec:elph} above, this stands in contrast to the case of scattering by phonons. 
Using (\ref{eq:MFL}) in the Kramers-Kronig integral (\ref{eq:overZ}) implies that at temperatures well below the cutoff
\be\label{eq:ZMFL}
\left. \frac{1}{Z} \right|_{T \ll T_\star} = 1 + \frac{2 \l}{\pi} \log \frac{T_\star}{T} + \cdots \,.
\ee
The $\cdots$ refer to non-universal terms, coming from frequencies around or above the cutoff, that do shift the mass but which are small compared to the singular term shown when $T \ll T_\star$. As we discuss further in \S\ref{sec:entropy} below, the residue (\ref{eq:ZMFL}) implies a logarithmic enhancement of the specific heat coefficient $c/T$ at low temperatures. It is indeed the case that in many quantum critical metals a low temperature $T$-linear resistivity is observed to coexist with a logarithmic specific heat. An illustrative example is shown in Fig.~\ref{fig:MFL}.
Other examples include the ruthenate Sr$_3$Ru$_2$O$_7$
\cite{Rost16549}, the cuprates Eu- and Nd-LSCO \cite{Michon2019, Daou2009} and several heavy fermion materials \cite{PhysRevLett.72.3262, PhysRevLett.85.626, PhysRevLett.91.257001}, although the analysis
of these particular cuprates and ruthenate are complicated by nearby van Hove-like features in the band structure that can also logarithmically enhance the specific heat. In such quantum critical phase diagrams, detuning from the critical point gaps the bosonic mode and hence introduces a {\it lower} cutoff $\Delta$ on the range of frequencies and temperatures over the which the MFL form (\ref{eq:MFL}) holds. At the critical point $\Delta \to 0$.

\begin{figure}[h]
    \centering
    \includegraphics[width=0.65\textwidth]{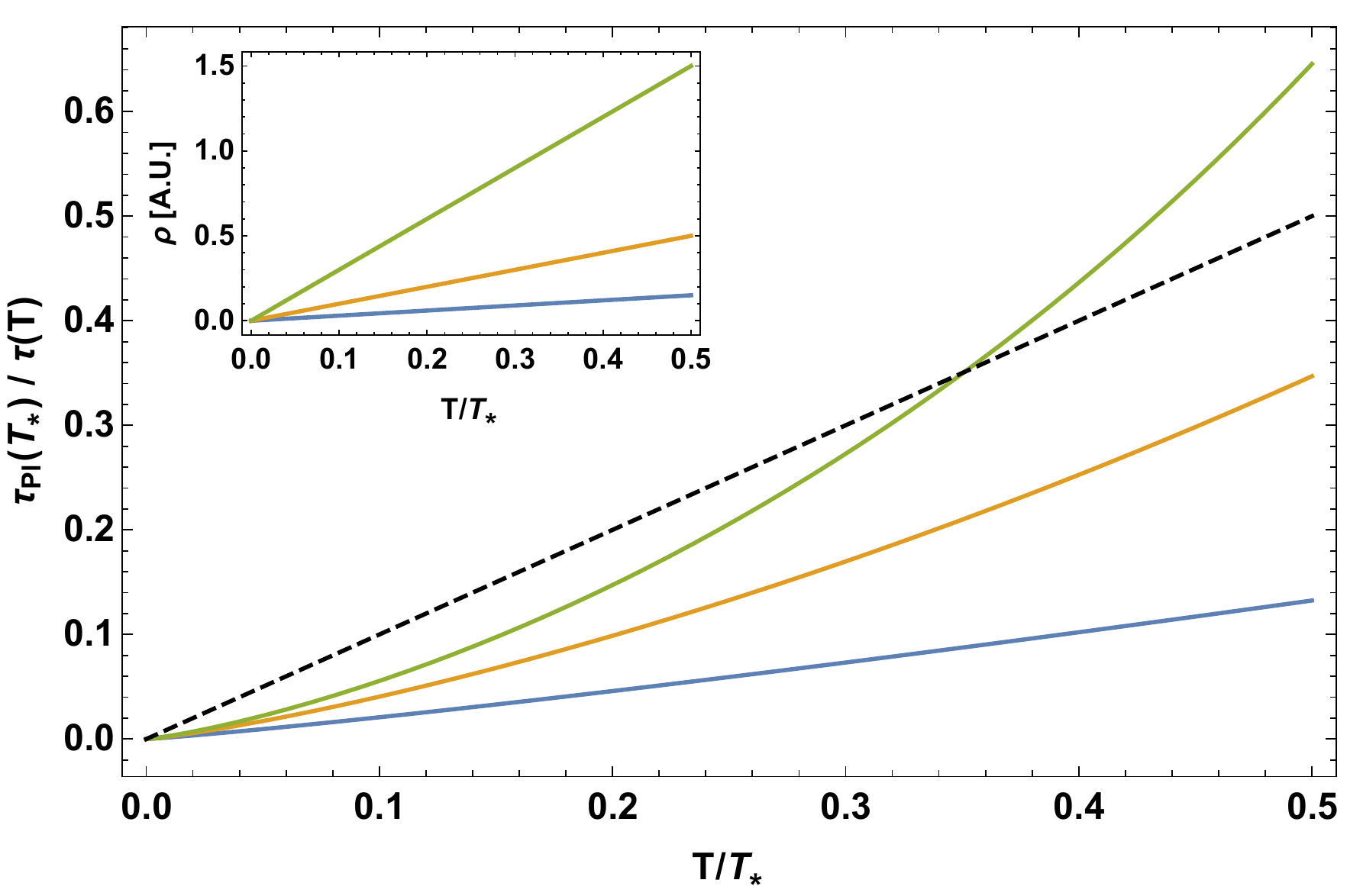}
    \includegraphics[width=0.288\textwidth]{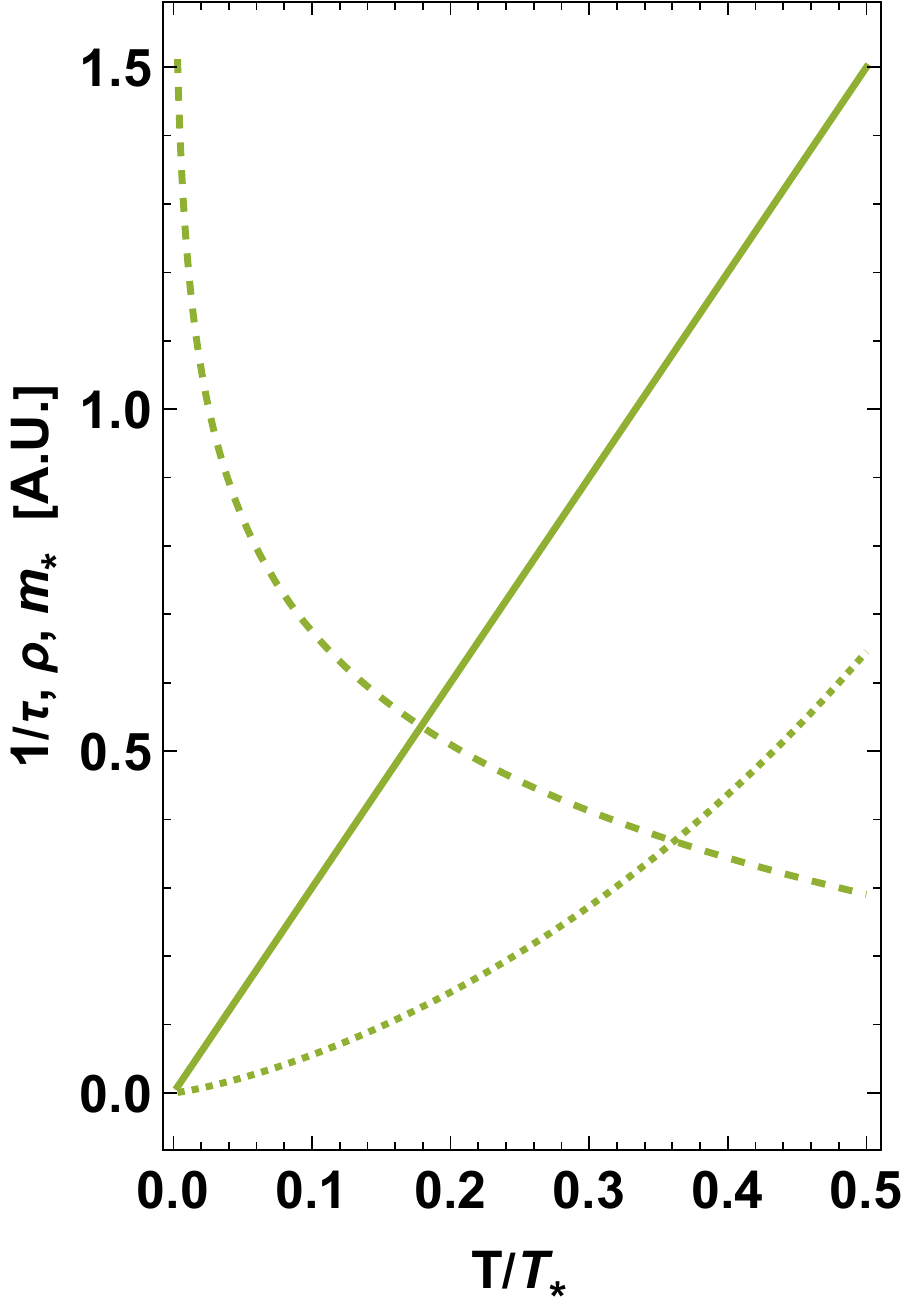}
    \caption{The MFL decay rate (\ref{eq:tauMFL}) as a function of temperature for $\lambda = 0.3, 1, 3$ (bottom to top). The dashed line shows the Planckian rate. At $T \ll T_\star$ the decay rate is sub-Planckian but, for large coupling $\lambda$, the Planckian limit can be surpassed as temperature is increased. The inset shows the corresponding $T$-linear resistivity, with slope proportional to $\lambda$. The right hand plot shows the temperature evolution of $1/\tau$ (dotted), $m_\star \propto 1/Z$ (dashed) as well as $\rho \propto m_\star/\tau$ (solid) for $\lambda=3$.}
    \label{fig:MFLrho}
\end{figure}

From (\ref{eq:MFL}) at $\omega = 0$ and (\ref{eq:ZMFL}), the physical lifetime (\ref{eq:tau}) is
\be\label{eq:tauMFL}
\left. \frac{1}{\tau} \right|_{T \ll T_\star} = \frac{\lambda}{1 + \frac{2 \l}{\pi} \log \frac{T_\star}{T}} \times \frac{k_B T}{\hbar}\, \leq \, \frac{1.57}{\log \frac{T_\star}{T}} \times \frac{k_B T}{\hbar} \,.
\ee
Therefore the physical decay rate of a MFL is Planckian bounded, in fact it obeys a logarithmically stronger bound, at low temperatures.\footnote{However, if $\Sigma'' = - \lambda \max(|\omega|,\a k_B T/\hbar)$ then the dimensionless number $\a$ will multiply the physical decay rate (\ref{eq:tauMFL}). In at least one microscopic realization of a MFL \cite{Mousatov2852}, it is the case that $\a \sim 1$.} The fact that the physical scattering rate is driven logarithmically small relative to the Planckian rate at low temperatures should help a Boltzmann description of transport.
A Planckian rate (up to a logarithm) is achieved
when the singular correction dominates $Z$:
at sufficiently low temperatures or for large $\lambda$. This may be a plausible origin for several of the observed Planckian scattering rates in strange metals, as discussed further in \S\ref{sec:transportime} below. We also recall in \S\ref{sec:transportime} below that the residue $Z$ does not appear in the resistivity, which therefore remains $T$-linear with no logarithmic correction. The decay rate (\ref{eq:tauMFL}) together with the corresponding resistivity is shown in Fig.~\ref{fig:MFLrho}.

As temperatures are increased up towards $T_\star$ the scattering rate (\ref{eq:tauMFL}) increases relative to the Planckian rate. That is, $\tau_\text{Pl}/\tau$ increases with temperature. In terms of a Planckian bound, then, the more dangerous regime is at high temperatures. This fact is illustrated in Fig.~\ref{fig:MFLrho}. Extrapolating (\ref{eq:tauMFL}) to estimate the value at the cutoff temperature $T_\star$ the inverse lifetime will be roughly the bare scattering rate
\be\label{eq:high}
\frac{1}{\tau(T_\star)}\approx - \Sigma_o'' = \l \frac{k_B T}{\hbar} \,.
\ee
In order for (\ref{eq:high}) to obey a Planckian bound the coupling itself must be bounded: $\lambda \lesssim 1$.\footnote{The description in terms of the original electronic quasiparticles may break down at these temperatures once $\lambda$ becomes large. This allows an alternative possibility that, as occurs in strong-weak coupling dualities, a different (`dual') slowly thermalizing description will emerge at large $\lambda$ in terms of new collective variables. Dualities are a clear instance where higher order processes, beyond simple scattering, are important to understand the reorganization of degrees of freedom and corresponding fundamental constraints on coupling constants. Furthermore, even without the emergence of dual quasiparticle degrees of freedom, higher order processes may lead to a saturation of the scattering cross section, as occurs for example in the low-T scattering of a unitary Fermi gas \cite{threview}.
}
Because a MFL has, presumably, a purely electronic origin it is plausible that the coupling is indeed subject to a many-body constraint. In the following section we outline a possible (and to our knowledge, new) entropic argument for such a bound on the coupling.

\subsection{Entropic scattering bound in a marginal Fermi liquid}
\label{sec:entropy}

In the course of preparing this article, we realized that if interactions are too strong over a wide temperature range, then they lead to an accumulation of entropy that is inconsistent with the microscopic Hilbert space of the system. This will be the case insofar as the electronic dynamics is captured by a lattice model with a finite `on-site' Hilbert space dimension, such as the Hubbard model. The strength of scattering, as measured by the imaginary part of the self-energy, is related via the Kramers-Kronig relation to the renormalized residue $Z$. When the momentum dependence of the self-energy can be neglected, this quantity in turn controls the renormalization of the Fermi velocity, and hence the density of states at the Fermi surface, $N_\star(0)$, and thereby the specific heat $c$. As an estimate
\be\label{eq:gamma}
\frac{c}{T} \sim k_B^2 N_\star(0) \sim  \frac{k_B^2 k_F^{d-1}}{\hbar v_F^\star} \propto \frac{1}{Z} \,.  
\ee
Here $d$ is the number of spatial dimensions.
In particular, then, according to the Kramers-Kronig relation (\ref{eq:overZ}), when the scattering is large the specific heat also becomes large.

The integral of the specific heat up to any finite cutoff $T_\star$ is bounded by the infinite temperature entropy density
\be\label{eq:boundS}
\int_0^{T_\star} \frac{c(T)}{T} dT \leq \frac{ k_B \log D}{a^d} \,.
\ee
Here $D$ is the dimension of the local on-site Hilbert space (e.g. $D=4$ if the states at any given site are no particle, spin up, spin down and two particles as in the Hubbard model) and 
$a^d$ denotes the volume of a lattice cell in $d$ dimensional real space.

The comments above are general, but in the remainder of this section we specialize to the case where a marginal Fermi liquid arises in such a lattice model below the temperature $T_\star$.
A MFL can be thought of as the minimal possible disruption to Fermi liquid theory. In particular, as we have discussed above, the MFL quasiparticle specific heat to be used in (\ref{eq:boundS}) only deviates logarithmically from the Sommerfeld behavior. This follows from the logarithmic form (\ref{eq:ZMFL}) of the residue for $T < T_\star$. We will neglect a possible additional contribution to the specific heat from the emergent collective bosonic mode responsible for the MFL scattering. Because such a contribution would `eat up' part of the microscopic entropy, including it would only strengthen the bound we will obtain on the quasiparticle specific heat.
Stronger non-Fermi liquid behavior can be expected to lead to stronger divergences of $c/T$ at low temperatures and hence potentially a stronger bound. To our knowledge this has not been observed.

The logarithmic temperature dependence of $c/T$ in a MFL does not make an important difference to the integrated specific heat. The main point is that if the MFL coupling $\lambda$ becomes large then the quasiparticle specific heat $c \propto \lambda$. With a large Fermi surface in two dimensions, i.e.~setting $k_F \sim 1/a$, the bound (\ref{eq:boundS}) is seen to imply that
\be\label{eq:lbound}
\lambda \lesssim \frac{T_F}{T_\star} \,.
\ee
Here $T_F \sim \hbar v_F k_F/k_B$ is the bare Fermi temperature. If the MFL behavior extends up to this scale, so that $T_\star \sim T_F$, then the bound requires $\lambda \lesssim 1$. As we have just discussed, this is the condition for the MFL to obey the Planckian bound at high temperatures. Super-Planckian scattering is allowed by this argument if the temperature range of strong scattering is significantly reduced so that $T_\star \ll T_F$. From this point of view, (\ref{eq:lbound}) can also be read as an upper bound on $T_\star$ given $\lambda$.\footnote{
Alternatively, the onset of a low entropy regime could cut off the integral in (\ref{eq:boundS}) at low temperatures. Note that ordering typically redistributes but does not reduce the total entropy below the ordering temperature, see e.g. \cite{RevModPhys.56.755, Petrovic_2001, Hartnoll:2012pp}.}

The entropic bound we have just described becomes weak in large $N$ theories, of the type discussed below in section \ref{sec:lyapunov}. Such models have many degrees of freedom per site, so that $D \sim N$ in (\ref{eq:boundS}). One may imagine that these models are effective theories where each site represents a corse-graining of many microscopic sites. Nonetheless, these effective theories miss microscopic entropy constraints of the kind we have described. This fact may deserve further consideration. The entropic argument we have given is also more difficult to apply to the electron-phonon problem. The phonon entropy diverges logarithmically at high temperatures and at intermediate temperatures may not be separable from the electronic entropy, especially if the electron-phonon coupling is large.
However, the general point that causality relates the real and imaginary parts of the self-energy opens the way, potentially, to other versions of entropy-driven bounds, something that we believe merits future attention.


\section{Transport lifetimes}
\label{sec:transportime}

Although the quasiparticle lifetime is in principle accessible from analysis of angle-resolved photoemission spectra \cite{RevModPhys.75.473, sunko2019angle}, photoemission experiments are restricted for practical reasons to a subset of the systems in which Planckian dissipation has been reported.    As we saw in \S\ref{sec:drude}, extensive experimental data instead comes from transport measurements, particularly those of dc transport, motivating a more detailed discussion of transport lifetimes.

Transport is concerned with the dynamics of conserved densities and currents. We will focus on electrical transport, for which the basic object is the dynamical conductivity $\sigma(\omega)$, which can be directly measured by optical experiments \cite{RevModPhys.83.471}. The $\omega \to 0$ limit is, of course, also accessed by dc transport measurements. In this section we consider in detail the differences between transport and quasiparticle lifetimes. It is important to be aware of the potential pitfalls of associating one too naively with the other.

Shining light onto a metal excites electron-hole pairs that will then dissipate. The conductivity is therefore a two-particle Green's function. The dissipation of an electron-hole pair is not, in general, reducible to that of the electron and the hole separately. For example, the electron and hole can interact via phonon exchange. The extent to which two-particle dynamics is not reducible to one-particle dynamics is captured by vertex corrections. The full dissipative dynamical conductivity can be written
\be\label{eq:opticalfull}
\text{Re} \, \sigma_{ij}(\omega) = \frac{4 e^2}{\hbar} \int \frac{d^dk}{(2\pi)^d} v_{Fi} \int \frac{d\Omega}{\pi} \Gamma_j(k,\Omega,\Omega+\omega) A(k,\Omega) A(k,\Omega+\omega) \frac{f(\Omega) - f(\Omega+\omega)}{\omega} \,.
\ee
The spectral weight $A(k,\Omega) = \text{Im} G^R(k,\Omega)$, using the full Green's function (\ref{eq:Green}). The electron and hole have relative momentum $2k$, relative energy $2\Omega$, and carry net energy $\omega$. The expression (\ref{eq:opticalfull}) is somewhat schematic; a more detailed description of the structure of the vertex function $\Gamma_j(k,\Omega,\Omega+\omega)$ can be found in e.g.~\S 8.1.2 and \S 8.4.2 of \cite{mahan}. The bare Fermi velocity $v_F$ appears explicitly in (\ref{eq:opticalfull}), from the microscopic definition of the current operator. The vertex function is seen to give an additional $k$- and $\Omega$-dependent weighting of the dissipative channels available to the electron-hole pair, amounting to a quantum distribution function for current-carrying electron-hole pairs.

In general the dynamical conductivity $\sigma(\omega)$ will be complicated and there will not be a unique timescale that can be extracted. We now describe assumptions that lead to a simple-minded Drude peak, whose width defines a transport lifetime. In particular, suppose that the processes dominating the conductivity integral (\ref{eq:opticalfull}) occur close to the Fermi surface. This assumption has two parts. Firstly, that the full Green's function can be expanded about the Fermi surface to give the simpler quasiparticle Green's function (\ref{eq:Green2}). This can be problematic. Even when the external frequency $\omega$ is very small, the electron frequency $\Omega$ can take values of order the temperature. We noted in our discussion of (\ref{eq:tau}) that this is large enough to invalidate the low frequency expansion of the Green's function for quantum critical electrons. Secondly, assume that the vertex function can be evaluated on the Fermi surface in the sense that $\Gamma_j(k,\Omega,\omega) \to \Gamma_j(k_F,\Omega,\omega)$. With these assumptions it is simple to perform the $k_\perp$ integral (orthogonal to the Fermi surface).\footnote{That is, 
\begin{align*}
\int_{-\infty}^\infty dk_\perp \frac{Z \tau}{1 + \t^2(\omega - v_F^\star k_\perp )^2}\frac{Z \tau}{1 + \t^2(\omega+\Omega- v_F^\star k_\perp)^2}
= \frac{\pi Z^2 \tau}{2 v_F^\star} \frac{1}{1 + (\t \omega/2)^2} \,.
\end{align*}
} To keep the formulae cleaner in this purely illustrative treatment, we will assume an isotropic Fermi surface. One obtains
\be\label{eq:opt2}
\text{Re} \, \sigma(\omega) = \frac{2 e^2}{d (2 \pi)^d}  \frac{A_\text{FS}}{\hbar v_F^\star} \frac{\tilde v^{\star \, 2}_{F} \, \Gamma(\omega) \, \tau}{1 +\left(\tau \omega/2 \right)^2} \,.
\ee
Here $A_\text{FS}$ is the area of the Fermi surface and we set $\tilde v_F^\star \equiv Z v_F = v_F^\star - Z \nabla_k \Sigma_o'$, see (\ref{eq:tau}) above; $\tilde v_F^\star$ is only equal to the renormalized Fermi velocity $v_F^\star$ when there is no $k_\perp$ dependence of the self-energy. The factor of two in the Drude peak denominator is because the electron and hole are both excited with half the energy of the frequency-dependent source. Finally, we define the vertex weighting $\Gamma(\omega) = \frac{1}{v_F} \int d\Omega \Gamma(k_F,\Omega,\Omega+\omega) [f(\Omega)-f(\Omega+\omega)]/\omega$.

Even with the above assumptions about dominance of near-Fermi surface physics, the function $\Gamma(\omega)$ in (\ref{eq:opt2}) is rather unconstrained in general. The Drude formula is obtained if the electron and hole are uncorrelated, so that $\Gamma(\omega) = 1$. Further assuming for simplicity that the $k_\perp$ dependence of the self-energy can be neglected, so that $\tilde v^\star_F = v^\star_F$ in (\ref{eq:opt2}), one obtains
\be\label{eq:opt3}
\text{Re} \, \sigma(\omega) = \frac{n e^2}{m_\star} \frac{\tau}{1 + \left(\tau \omega/2 \right)^2 } \,.
\ee
The mass $m_\star = \hbar k_F/v_F^{\star}$ and the electron density $n \propto k_F^d$.
The most immediate conclusion from (\ref{eq:opt3}) is that, unsurprisingly, it is the renormalized, physical decay rate $1/\tau$ that sets the width of the Drude peak (and not the bare quantity $\Sigma''_o)$.
In this simplified model, isotropic and without vertex corrections, the peak in the dynamical conductivity is twice the width of the single-particle peak. The $\omega \to 0$ limit of (\ref{eq:opt3}) is of course\footnote{We reintroduced the density $n$ in (\ref{eq:opt3}) to connect to the textbook Drude formula (\ref{eq:dc}). However, it should be emphasized that all low energy scattering takes place at the Fermi surface. It is therefore more physically transparent to write the dc conductivity as $\sigma = \frac{1}{d} e^2 v_F^{\star\,2} N_\star(0) \tau$, with $N_\star(0)$ the renormalized density of states at the Fermi surface.}
\be\label{eq:dc}
\sigma = \frac{n e^2 \tau}{m_\star} \,.
\ee
The dc conductivity can equivalently be written --- in the highly simplified setting currently being discussed --- in terms of unrenormalized quantities as $\sigma = n e^2/(m |\Sigma_o''|)$, with the factors of $Z$ cancelling out. The renormalized mass $m_\star$ is the one measured by quantum oscillations and the specific heat (insofar as these measure single-particle properties). Using the physical mass $m_\star$ in a dc Drude analysis will extract the physical timescale $\tau$. This procedure is widely used in determining the $T^2$ scattering rate in Fermi liquids, as discussed in Appendix \ref{sec:KW}.

Correlations between the electron and hole, captured by $\Gamma(\omega)$, are important when there are scattering processes that can strongly degrade a given single-particle Bloch state without significantly degrading the current. For example, small-angle scattering does not efficiently degrade current. A second example is that umklapp scattering, as opposed to momentum-conserving normal scattering, is necessary in order to fully degrade the current as long as the metal is not compensated, i.e. does not have equal densities of electrons and holes. The role of vertex corrections for small-angle scattering is discussed in e.g.~\S8.4.2 of \cite{mahan} and for umklapp scattering in e.g.~\cite{Maslov_2016}.\footnote{In \S\ref{sec:qc} we noted that quantum criticality commonly leads to small angle scattering. Furthermore, quantum critical scattering is often momentum-conserving, so that the effective critical theory has an emergent conserved momentum and an infinite conductivity. Non-critical umklapp processes (or disorder) must then be incorporated, again weakening the link between criticality and transport \cite{PhysRevLett.106.106403,pal2012resistivity}.
A powerful technique for incorporating momentum relaxation is the memory matrix formalism \cite{PhysRevB.75.245104, PhysRevLett.108.241601, PhysRevB.89.155130}.
} The vertex weighting function $\Gamma(\omega)$ essentially counts the fraction of scattering processes that degrade current when the electron-hole pair has total energy $\omega$. In the simplest circumstances, this effect amounts to a shift
\be\label{eq:tautr}
\tau \to \tau_\text{tr} \,,
\ee
in the dynamical and dc conductivities (\ref{eq:opt3}) and (\ref{eq:dc}). A well-known instance of the difference between the single-particle lifetime $\tau$ and the transport lifetime $\tau_\text{tr}$ is for scattering by phonons at low temperatures, where the scattering rate $T^3 \to T^5$. We saw in \S\ref{sec:qc} that quantum critical metals with scattering by long-wavelength critical bosons have an analogous shift $T \to T^{1+2/z}$, with $z$ the dynamic critical exponent (and a MFL has $z=\infty$).

Even when, after allowing for the shift (\ref{eq:tautr}) in the lifetime, the dynamical conductivity retains the Lorentzian form (\ref{eq:opt3}), the vertex function has every right to change the weight of the Lorentzian (the `Drude weight'). That is,
the effective mass $m_\star$ that appears in the Drude formula need not be the single-particle renormalized mass. One cannot therefore, in general, extract $\tau_\text{tr}$ from dc measurements alone.
While this vertex-induced renormalization of the Drude weight has been argued to be small when the electronic self-energy has only weak momentum dependence, i.e.~is local, recent numerical results on the Hubbard model at high temperatures found otherwise \cite{PhysRevLett.123.036601}.

In summary, there are many obstacles to the existence of an unambiguously defined, unique transport lifetime. We had already noted in the previous \S\ref{sec:single} that single-particle lifetimes are not well defined if the self-energy depends strongly on frequency. In this section we have seen that even if the single-particle lifetime is well defined, the transport lifetime may not be if the vertex function --- which is an eminently physical quantity --- is nontrivial. And finally, even if the dynamical conductivity is Lorentzian, so that there is a well defined transport lifetime, then in general this cannot be determined from dc measurements because there will be an unknown renormalization of the Drude weight.

These difficulties suggest that, in the absence of an understanding of what is causing the Planckian behavior of many unconventional metals, the Drude analyses summarized in \S\ref{sec:drude} should be interpreted as an informed dimensional analysis in which measured quantities $\{\sigma, n, m_\star\}$ are used to define a timescale. This timescale is then taken to be a proxy for the equilibration time. 
It may be that, a posteriori, it is understood that the underlying scattering in these systems is something relatively simple (e.g.~phonon-like\footnote{It need not literally be phonons that provide the scattering. For example, there may be circumstances where spins can play a similar role.} or some version of a MFL). The robustness of the Planckian lifetime extracted from the Drude analyses in \S\ref{sec:drude} may perhaps be taken as indirect evidence for a simple underlying mechanism. In this case, it may be possible to elevate these analyses to a precision science.

\subsection{Drude analyses revisited: value of the effective mass}
\label{sec:drude2}

With all the caveats of the above sections in mind, here we
discuss several more prosaic issues that are pertinent to the Drude analyses of \S\ref{sec:drude}. These are primarily related to the value of the effective mass that should be used.

To estimate a timescale $\tau$ using the Drude formula (\ref{eq:dc}) measurements are needed of the resistivity, effective mass $m_\star$ and density $n$. The resistivity is unambiguous. The density is also straightforward, in principle at least. The Luttinger count fixes the density in terms of the number of itinerant carriers, which may be known a priori. Quantum oscillations provide a direct experimental probe of the density. The frequency of the oscillations is determined by the cross-sectional area of the Fermi surface \cite{onsager}. In two dimensions summing the areas of the observed Fermi surface sheets directly gives the density. While the Hall coefficient is sometimes used as a proxy for the density, in the weak field limit that is typically accessible it is sensitive to both the nature of scattering and the Fermi surface geometry, as in e.g. \cite{PhysRevB.43.193}, and furthermore cannot account for the simultaneous presence of both electrons and holes.

The effective mass is more complicated because it can vary around the Fermi surface and between different sheets. This fact becomes significant if the Fermi surface develops large mass `hot spots' or `hot pockets' close to a quantum critical point. Two commonly used probes to estimate the effective mass are quantum oscillations and the specific heat. The amplitude of quantum oscillations depends on the cyclotron mass \cite{lifshitz1956theory} while the specific heat depends on the density of states at the Fermi surface, see (\ref{eq:gamma}) above. Both of these quantities are sensitive to an averaged mass around the Fermi surface $\langle m_\star \rangle_\text{FS}$ while dc transport instead depends on the average of the inverse mass $\langle m^{-1}_\star \rangle_\text{FS}$, if the timescale $\tau$ is kept uniform. In particular, heavy regions of the Fermi surface are not important for transport but can potentially dominate these probes.

Quantum oscillations have the advantage of separating out the contributions from distinct Fermi sheets --- these correspond to different oscillation frequencies. 
This is especially helpful in cases, such as Sr$_3$Ru$_2$O$_7$, where a single sheet develops a large density of states due to incipient quantum criticality \cite{PhysRevLett.101.026407, PhysRevB.81.235103}. The carriers in the heavy sheet do not contribute significantly to transport but can dominate the specific heat \cite{Rost16549, Mousatov2852}. In these circumstances the total specific heat will not extract the correct averaged mass for use in the Drude formula. Quantum oscillations will instead determine sheet-wise masses $m_{\star\, i}$ and densities $n_i$ that can be used to define a timescale \cite{Bruin804}
\be\label{eq:ave}
\frac{1}{\tau} = \frac{e^2}{\sigma} \sum_i \frac{n_i}{m_{\star\,i}} \,.
\ee
In this formula heavy sheets --- which might not even be detected by quantum oscillations --- are correctly seen to give a suppressed contribution to the transport lifetime. As we noted in \S\ref{sec:drude} above, the expression (\ref{eq:ave}) involves a further approximation in which a unique timescale is associated to the entire Fermi surface. This level of averaging is unavoidable in the Drude approach, and is less likely to introduce large errors to the analysis than a procedure in which an overall average mass is obtained from the heat capacity.  In heavy fermion compounds (which in reality are heavy-and-light fermion compounds), for example, the estimate involving only the heat capacity can be a source of serious errors.

An effective mass can also be found from analyses of optical data for $\sigma(\omega)$, as it enters the Drude weight in e.g.~(\ref{eq:opt3}). Such an analysis will simultaneously determine a transport timescale. Even if the Drude peak is not Lorentzian, estimation of a timescale and weight from optical data offers an opportunity to cross-check the values of both quantities obtained from the dc Drude analysis. It is important, however, to allow for the possibility that the optical data should be analyzed as a sum of Drude peaks, especially in multiband compounds (see e.g.~\cite{PhysRevB.103.205109} for a two-Drude analysis of the $T^2$ resistivity regime of several ruthenates).

\begin{figure}[h]
    \centering
    \includegraphics[width=0.7\textwidth]{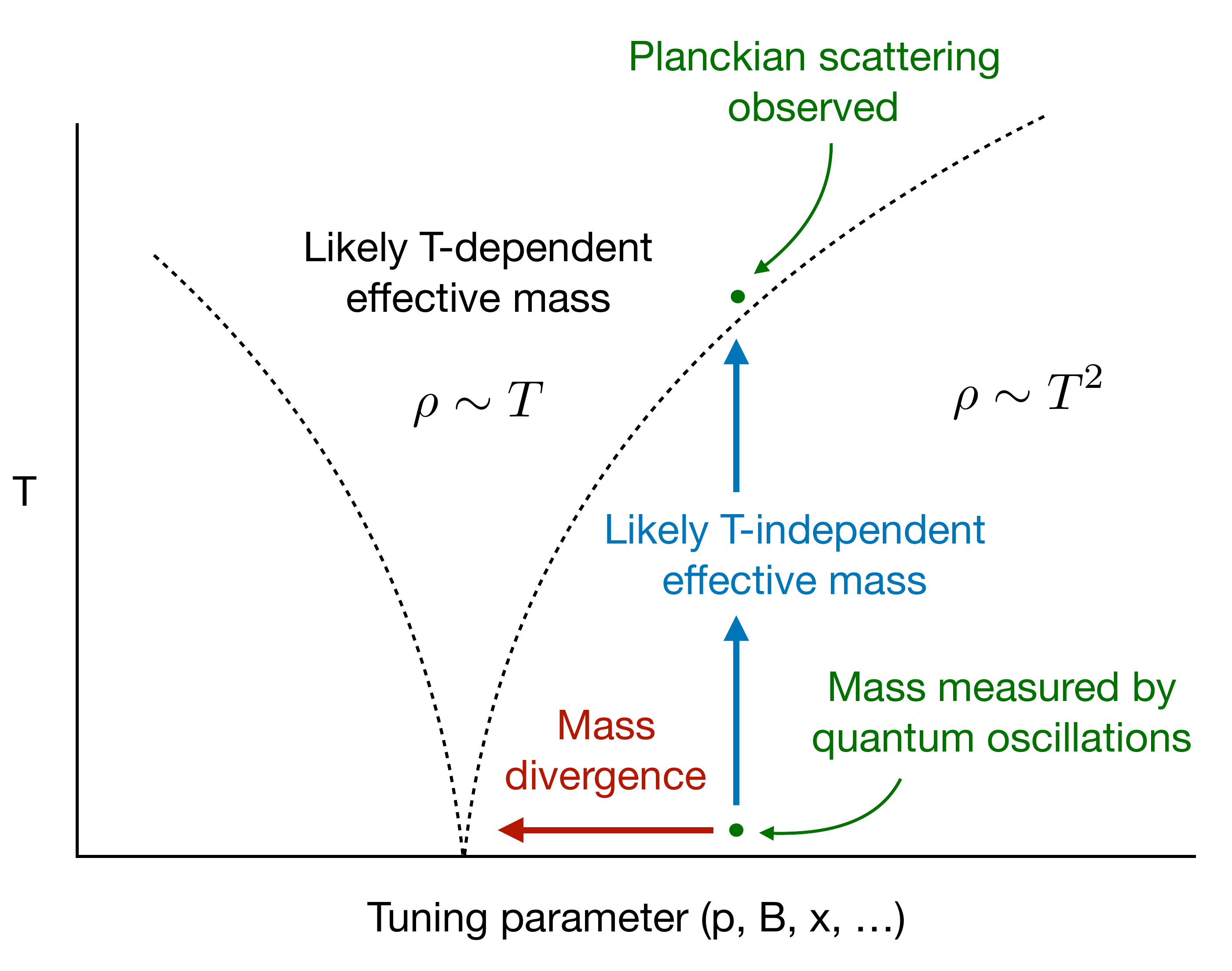}
    \caption{Schematic diagram showing the evolution of the effective mass as a function of temperature and tuning parameter. Quantum oscillations measure the low temperature mass, while Planckian scattering is claimed at the lower end of the $T$-linear transport regime. Note that observed mass divergences as the critical point is approached at low temperatures do not necessarily correspond to the quasiparticles that dominate transport.}
    \label{fig:masses}
\end{figure}

Even if sheet-by-sheet carrier densities and masses can be obtained from quantum oscillations, there is another potential concern. The amplitude of quantum oscillations decays exponentially with temperature and with the effective mass $m_\star$. For this reason the values of the mass and density used in the Drude analysis of e.g. \cite{Bruin804} and \cite{Licciardello2019} are extracted from a low temperature Fermi liquid-like regime with a $T^2$ resistivity. In contrast, $T$-linear resistivity onsets at temperatures $T>T_\Delta \equiv \Delta/k_B$, for some energy scale $\Delta$. The objective is to obtain the transport timescale for this higher temperature regime. One must worry, therefore, about the possible temperature dependence of the Drude weight $n e^2/m_\star$. The more plausible source of temperature dependence is the effective mass. There are two potential issues: temperature dependence of $m_\star$ in the low temperature (putatively Fermi liquid) $T^2$ regime and temperature dependence in the high temperature $T$-linear regime. We consider these in turn, with the discussion summarized in Fig.~\ref{fig:masses}.

In a Fermi liquid the imaginary part of the self-energy takes an especially simple form $\Sigma_\text{FL}'' \propto (\pi k_B T)^2 + (\hbar \omega)^2$, see e.g.~\cite{PhysRevB.86.155136}. This means that the quantity that appears in the Kramers-Kronig integral (\ref{eq:overZ}) for the effective mass, $\Sigma_\text{FL}''(\omega) - \Sigma_\text{FL}''(0) \propto \omega^2$, is indeed independent of temperature. However, if $T$-linear scattering onsets above $T_\Delta$ then $\Sigma''$ will differ from $\Sigma_\text{FL}''$ above that temperature. It is likely, then, that its frequency dependence will also differ at frequencies greater than $\Delta/\hbar$ (for example, this would be the case for a MFL). Assuming that $\Sigma'' = \Sigma_\text{FL}''$ for $\max(\hbar \omega,k_B T) < \Delta$ one obtains that the change in the effective mass over the Fermi liquid temperature range is
\be\label{eq:mT}
m_\star(T_\Delta) - m_\star(0) = - \frac{1}{\pi} \int_{\Delta/\hbar}^{\omega_\star} \frac{\Sigma''(\omega,T_\Delta) - \Sigma_\text{FL}''(\omega,T_\Delta)}{\w^2} d\omega \,.
\ee
Here $\omega_\star$ is some high-frequency cutoff, for example as introduced in \S\ref{sec:MFL}. The temperature dependence implied by (\ref{eq:mT}) will be weak if the full scattering rate $\Sigma''$ only starts to differ significantly from $\Sigma_\text{FL}''$ at frequencies somewhat above $\Delta/\hbar$, so that the difference $\Sigma'' -\Sigma_\text{FL}''$ is suppressed by the factor of $1/\omega^2$ in (\ref{eq:mT}). There is no evidence for a temperature dependence of the mass in specific heat data in low temperature Fermi liquid regimes, e.g. \cite{AOKI1998271, PhysRevLett.91.257001, Rost16549}, although the presence of a phonon contribution to the specific heat can complicate the analysis. It may be interesting to understand this absence of temperature dependence of the effective mass in the light of (\ref{eq:mT}), and to search for possible temperature dependence using probes such as the Knight shift that do not have a phonon contribution. In addition to introducing temperature dependence, the high frequency modes renormalize the low temperature mass $m_\star(0)$. This low temperature mass obeys the Kadowaki-Woods relation, as we discuss in Appendix \ref{sec:KW}.

For $T > T_\Delta$, i.e.~in the $T$-linear regime, there is every reason to expect the effective mass to be temperature dependent. We saw in \S\ref{sec:single} that this was the case for both scattering by phonons and in a MFL. Strong temperature dependence of the effective mass in $T$-linear transport regimes has also been found in Dynamical Mean Field Theory (DMFT) studies of correlated Hubbard model-like systems \cite{PhysRevLett.110.086401, PhysRevLett.111.036401, PhysRevLett.113.246404, PhysRevLett.116.256401}. Similarly to the case of a MFL, as was noted in \S\ref{sec:MFL} above, the temperature-dependent mass in the DMFT studies helps to restore a quasiparticle description. If there is indeed no temperature dependence in the mass for $T < T_\Delta$ then, by continuity, the low temperature mass $m_\star(0)$ extracted from quantum oscillations should correspond to the mass just inside the lower end of the $T$-linear regime. This is shown in Fig.~\ref{fig:masses}.\footnote{This continuity argument assumes that strong mass renormalization does not onset in the Fermi liquid regime immediately below $T_\Delta$. Especially in a MFL, where mass renormalization is only logarithmic, this would seem to be a reasonable assumption.} We have seen that in simple models for both conventional (\S \ref{sec:phdeb}) and unconventional (\S \ref{sec:MFL}) $T$-linear transport the physical lifetime is automatically Planckian bounded at this lower temperature end. These (or similar) models may therefore offer the simplest explanation for many of the experimentally reported Planckian lifetimes.

As discussed in \S\ref{sec:single}, the mass renormalization in simple models of $T$-linear quasiparticle lifetimes becomes weaker at higher temperatures. The slope of the resistivity does not change, which means that if such simple models are correct the physical scattering rate is becoming stronger at high temperatures. Given that the scattering rate is already estimated, per the discussion immediately above, to be Planckian at the lower end of the $T$-linear temperature range, there is the distinct possibility of super-Planckian scattering at high temperatures. To probe this possibility directly it will be necessary to measure the effective mass (or otherwise extract a lifetime) at high temperatures in the $T$-linear transport regime. Although photoemission data already exist in the relevant regimes for materials such as cuprates \cite{RevModPhys.93.025006}, no such effects have, to our knowledge been convincingly observed to date.

A dc Drude analysis at high temperatures in
niobium-doped strontium titanate --- using a high temperature effective mass obtained from the thermopower, relative to a low temperature mass determined by quantum oscillations --- led to a claim of strongly super-Planckian decay rates at very high temperatures \cite{PhysRevX.10.031025}. A recent re-analysis, however, suggests that timescales do not in fact become super-Planckian \cite{nazaryan2021conductivity}. In any case, the electrons are non-degenerate at these high temperatures and the estimated decay rate goes as $1/\tau \sim T^{2.5}$. Given the unusual temperature dependence of many quantities, the nature of charge transport in this interesting regime needs to be understood better before reliable conclusions can be drawn about the meaning of timescales extracted from conventional formulae.

\subsection{Relation of $T$-linear resistivity to more general transport behavior}

Our discussion has mostly focused on systems exhibiting $T$-linear resistivity at low temperatures. These are plausibly the simplest unconventional metals, wherein quasiparticle concepts such as an effective mass may continue to be useful. To establish the existence of truly universal bounds on dissipation it will be necessary, ultimately, to grapple with strongly incoherent regimes of transport. These include heavy fermion systems at higher temperatures, underdoped cuprates and other systems close to Mott transitions. A `Planckian' analysis of transport in these cases will require a careful identification of the effective density of charge carriers and possibly a detailed understanding of the transport mechanism.

\begin{figure}[h]
    \centering
    \includegraphics[width=0.475\textwidth]{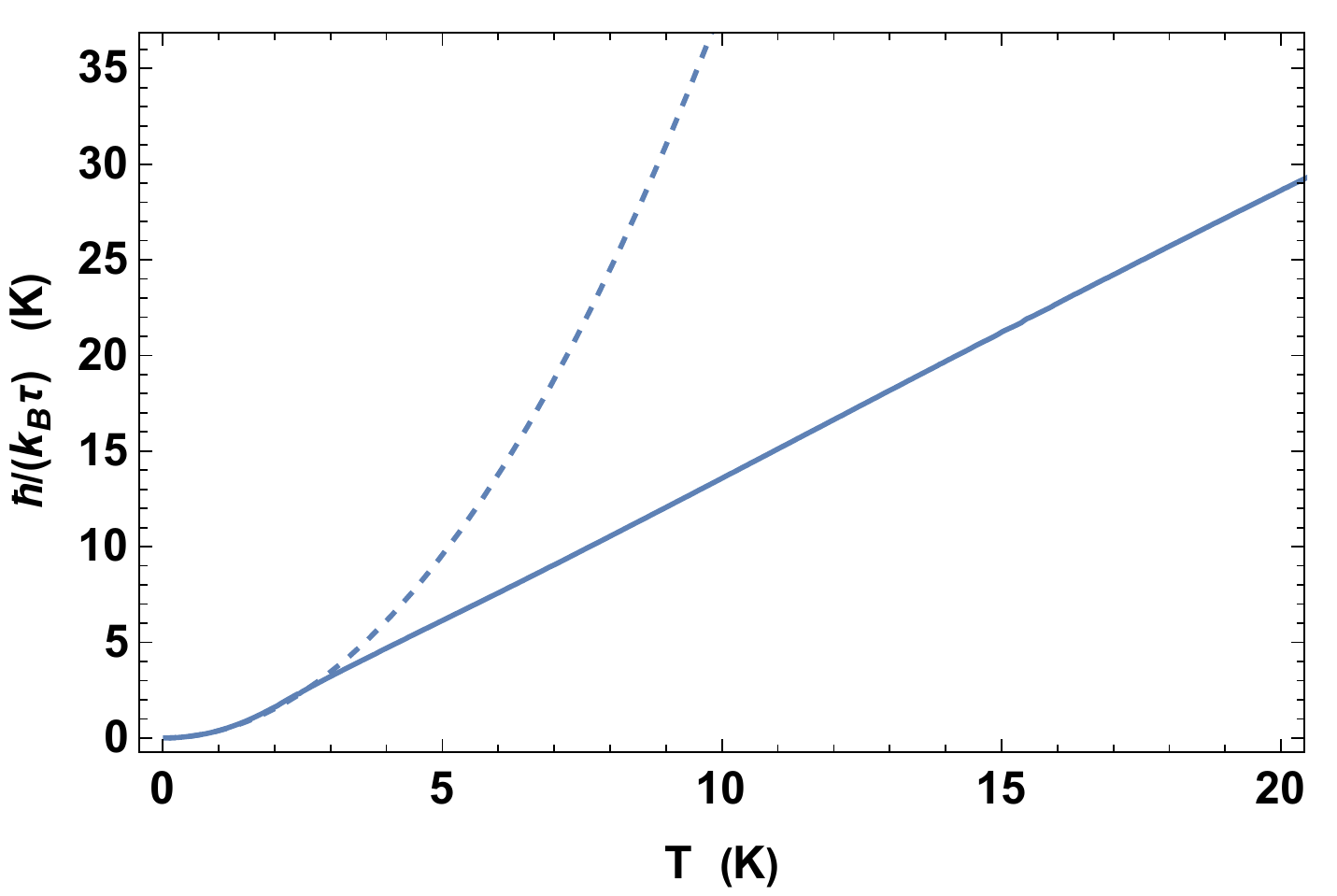}\includegraphics[width=0.49\textwidth]{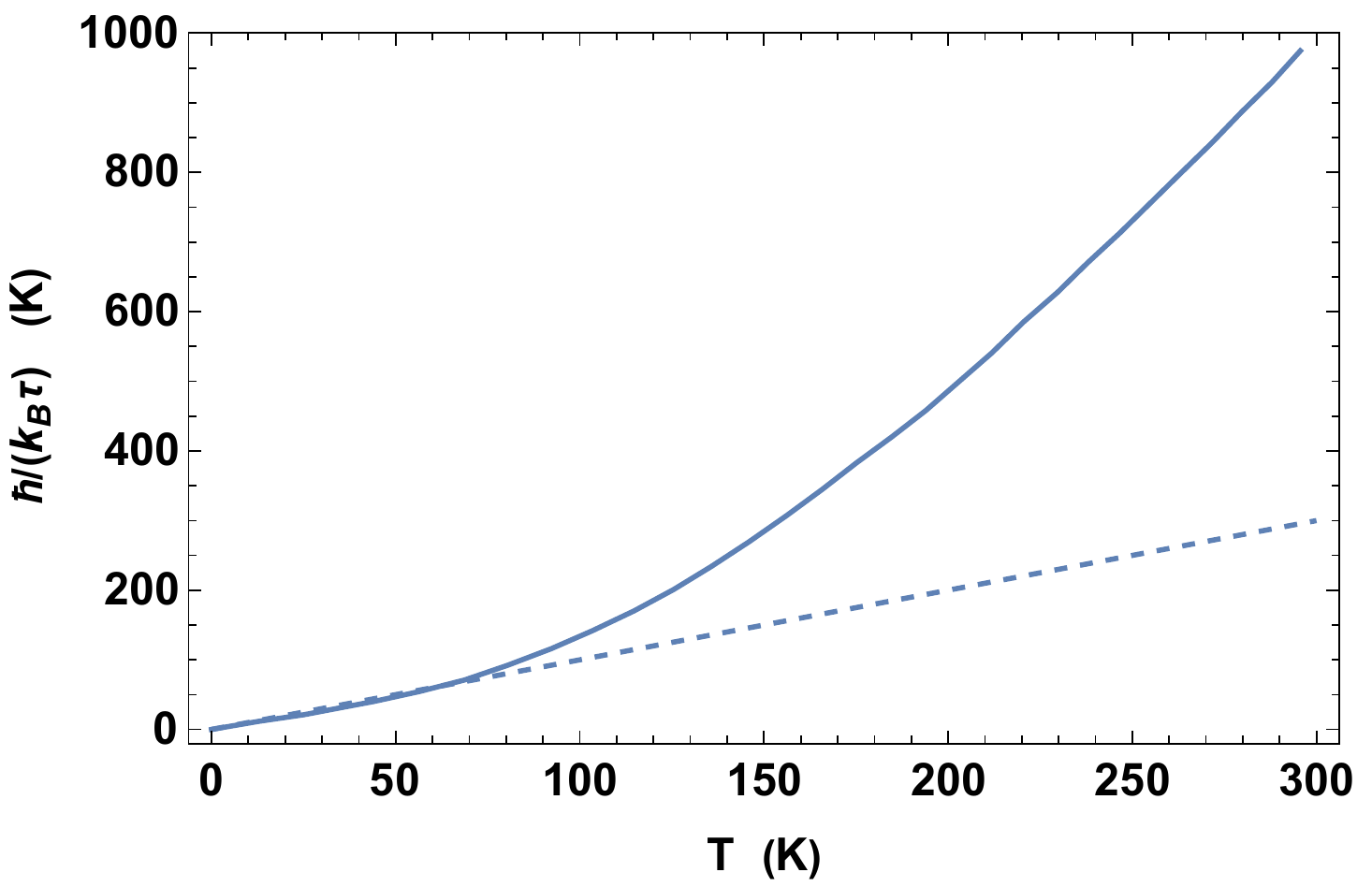}
    \caption{Left plot: a low temperature $T^2$ scattering rate crosses over to a high temperature $T$-linear scattering rate in Sr$_3$Ru$_2$O$_7$ at field $B=7.4$T \cite{Bruin804}. The solid line is data and the dashed line is the extrapolation of the low temperature behavior. Here the strong low temperature growth of the scattering rate is cut off by Planckian scattering. Right plot: a low temperature $T$-linear scattering rate \cite{Legros2019} crosses over to a high temperature $T^2$ scattering rate \cite{poniatowski2021counterexample} in LCCO at $x=0.16$. Here, the high temperature scattering rate rises above the low temperature Planckian scattering. The $T=0$ scattering rate has been subtracted in both cases. For Sr$_3$Ru$_2$O$_7$ the masses and densities used to extract a scattering rate from transport have been obtained from quantum oscillations \cite{Bruin804}. In the absence of quantum oscillation data on LCCO, the masses were estimated from quantum oscillations in NCCO \cite{Legros2019}. This procedure introduces some uncertainty in the density.}
    \label{fig:TT2}
\end{figure}

A more cautious step beyond pure $T$-linear resistivity are situations where $T$-linearity occurs in conjunction with a Fermi liquid-like $T^2$ behavior. Indeed, one of the original motivations for a Planckian bound was the observation that in many of the materials with phase diagrams like that sketched in Fig.~\ref{fig:masses}, the resistivity in the $T$-linear regime is {\it lower} than the extrapolation of the $T^2$ term to these temperatures \cite{Bruin804}. In such cases the $T$-linear resistivity is observed to be Planckian, a phenomenology naturally associated with a bound. This behavior is illustrated in Fig.~\ref{fig:TT2}, left plot.

In other cases, however, the resistivity takes the additive form $\rho \sim T + T^2$. As with the purely $T$-linear regimes, this behavior can arise over different temperature ranges. For example, in overdoped cuprates it is seen down to low temperatures \cite{PhysRevB.53.5848,Cooper603}, while in the ruthenate Sr$_2$RuO$_4$ it emerges above a low temperature Fermi liquid regime with purely $T^2$ resistivity \cite{PhysRevB.58.R10107}. The first point to make is that the simplest framework for understanding such behavior is a current carried by well-defined quasiparticles that undergo two additive scattering mechanisms. The second point is that there may seem to be some tension between the idea that the $T$-linear scattering is maximally dissipative and the fact that it is possible to `add' to it. A closely related objection is made in \cite{doi:10.1146/annurev-conmatphys-031119-050558}, and subsequently in \cite{poniatowski2021counterexample} it has been claimed that an overdoped electron-doped cuprate has $\tau_\text{Pl}/\tau \sim 4$ at room temperature. We show this data in Fig.~\ref{fig:TT2}, right plot. For that system and the others mentioned in this paragraph, however, the $T$-linear and $T^2$ components have comparable magnitudes over the temperature ranges in question and there does not appear to be any strongly super-Planckian scattering. Looking forwards, these may be especially interesting regimes in which to tease apart some of the different notions we have discussed throughout this review relating to Planckian bounds: the need for careful low temperature characterization of the effective mass using quantum oscillations, the possible temperature dependence of the mass and the distinction between elastic and inelastic contributions to scattering.

\section{Quantum Lyapunov exponent and butterfly velocity}
\label{sec:lyapunov}

The only rigorous Planckian bound that has been proven to date is on the quantum Lyapunov exponent $\lambda_L$ \cite{Maldacena:2015waa}, that characterizes the onset of many-body quantum scrambling. As we explain shortly, this exponent is probably not directly related to transport or equilibration in the experimental systems we have described above. Nonetheless, the bound on $\lambda_L$ is an important proof of principle success for the program of establishing fundamental bounds on many-body quantum dynamics. Furthermore, the quantum Lyapunov bound has driven the (re-)emergence of SYK models \cite{PhysRevLett.70.3339, kitaev2015talk, Maldacena:2016hyu} as a class of controlled theories of non-quasiparticle transport (relevant work predating the current interest includes \cite{PhysRevB.59.5341}).

This section will, by necessity, not be self-contained. We will discuss how the concepts of Lyapunov exponent and butterfly velocity (neither of which we will define explicitly) may be relevant to Planckian dissipation and how holographic and SYK models give theoretically controlled non-quasiparticle realizations of various ideas that have appeared throughout this review. For more extended reviews of these topics see \cite{Liu:2020rrn, Chowdhury:2021qpy}.

The quantum Lyapunov bound states that \cite{Maldacena:2015waa}
\be\label{eq:MSS}
\lambda_L \equiv \frac{1}{\tau_L} \leq \frac{2 \pi}{\tau_\text{Pl}}  \,.
\ee
In essence, this is a bound on how quickly an operator can grow under Heisenberg time evolution. Operator growth is a quantum-mechanical measure of the dynamical scrambling of information in phase space. As discussed in \cite{PhysRevB.98.144304}, see also references therein, this bounded timescale is only defined for large $N$ (or otherwise semiclassical) theories because for an operator to exhibit well-defined exponential growth it must have a large enough `on-site' phase space to grow into. This limits the direct applicability of the result to many interesting condensed matter systems which have small on-site Hilbert spaces. This limitation can be evaded, however, if operators are able to grow sufficiently rapidly in space, as we now discuss.

In addition to on-site growth, operators can also grow in space. In systems with spatially local microscopic interactions, the butterfly velocity $v_B$ \cite{Shenker:2013pqa, Roberts:2014isa} defines a light-cone that bounds the speed of operator growth in space \cite{PhysRevB.98.144304}. A sufficiently large butterfly velocity can allow exponential operator growth
even in systems with a small on-site Hilbert space \cite{PhysRevB.103.L121111}. It has been argued that the butterfly velocity is a temperature-dependent refinement of the more microscopic Lieb-Robinson velocity and hence a more physically relevant velocity at low temperatures \cite{Roberts:2016wdl}. Despite its name referring to the `butterfly effect', $v_B$ is formally nonzero in non-interacting systems and furthermore the operator growth light-front has the same form in chaotic and certain non-chaotic `interacting' integrable spin chains \cite{PhysRevB.98.220303}.

It was shown in \cite{Blake:2016wvh, Blake:2016sud}
that in various strongly non-quasiparticle and Planckian holographic models the diffusivity has the form $D = \a v_B^2 \tau_\text{Pl}$, with $\a$ a constant that does not depend strongly on microscopic data. This fact supported the idea that the butterfly velocity defines a physically relevant non-quasiparticle velocity.
Further works established that the butterfly velocity specifically controls heat diffusion in holographic \cite{Blake:2016jnn, Blake:2017qgd, Baggioli:2016pia} and SYK \cite{Gu:2016oyy, PhysRevB.95.155131, PhysRevB.100.045140} models of Planckian transport.
It has been argued \cite{Blake:2017ris, Blake:2021wqj} that this relationship can arise as a consequence of these models being `maximally chaotic', i.e.~saturating the bound (\ref{eq:MSS}). In maximally chaotic theories the energy dynamics is intimately tied up with quantum chaos. The importance of the holographic and SYK-type models, then, is that they bring out simplifications in quantum dynamics that arise in a non-quasiparticle Planckian limit that is strictly the opposite of the conventional weakly interacting quasiparticle regime. Various other classes of large $N$ models have also exhibited a close connection between energy diffusion and the butterfly velocity, e.g. \cite{Patel1844, PhysRevX.7.031047, werman2017quantum}.
This connection can be broken if a small number of fast modes dominate heat transport but inherit chaotic properties from a strongly interacting large $N$ bath \cite{PhysRevB.104.195113}.

Several of the works mentioned in the previous paragraph found that energy diffusion was related to the Lyapunov time as $D = v_B^2 \tau_L$. However, inhomogeneous SYK chains offer a counterexample to this connection \cite{Gu:2017ohj}.
In Appendix \ref{sec:diff} we argue that the equilibration time is the more natural timescale to control transport. While both $\tau_\text{eq}$ and $\tau_L$ are Planckian in the simplest holographic and SYK models, they can also be distinct \cite{Hartman:2017hhp, PhysRevLett.123.141601}. In \cite{PhysRevLett.122.216601, 10.21468/SciPostPhys.9.5.071} a picture of equilibration as operator growth was developed, straddling both timescales.

A systemic overview of the large and developing literature on SYK models of non-Fermi liquids is beyond the scope of our discussion here. We simply note some points of connection between this literature and the subject of Planckian bounds.
In \cite{PhysRevLett.119.216601} a heavy Fermi liquid emerges at low temperatures from an incoherent $T$-linear metal. We discuss the relation of strong scattering regimes to low temperature mass renormalization in \S\ref{sec:single} and \S\ref{sec:drude2} and, in particular, the associated buildup up of entropy in \S\ref{sec:entropy}. In \cite{PhysRevB.100.045140} it is shown that $T$-linear transport in the incoherent regime is related to strong operator growth while transport in $T$-linear regimes due to phonon scattering is not. This is due to the disjunction between equilibration and transport in high temperature elastic regimes of phonon scattering discussed in \S\ref{sec:elas} (cf.~also \cite{PhysRevLett.122.216601}).
In \cite{PhysRevX.8.031024} various non-Fermi liquid regimes are found in which the mass renormalization enforces a Planckian bound on physical transport lifetimes, as in our discussion in \S\ref{sec:phdeb} and \S\ref{sec:MFL} above. Robustly Planckian physical lifetimes are obtained in \cite{PhysRevLett.123.066601} from a model with a smeared out Fermi surface, whose similarities to a phenomenological `flat band' theory in which Planckian dissipation has also been analyzed \cite{Shaginyan2019} have been noted \cite{Volovik2019}.
Controlled theories of a Planckian marginal Fermi liquid, considered in \S\ref{sec:MFL}, were obtained in \cite{aldape2020solvable, esterlis2021large} by performing certain averages on a large $N$ model of fermionic quantum criticality. A marginal Fermi liquid was also obtained in \cite{Cha18341} at a critical point describing the quantum melting of a spin glass phase into a Fermi liquid.
In \cite{PhysRevResearch.2.033431} super-Planckian scattering in a model of strongly anharmonic phonons is avoided by the onset of a glass phase, possibly giving an instance of a `quantum engineering' constraint of the kind we mentioned in footnote \ref{foot:pol}. In \cite{PhysRevResearch.2.033434} a model is given exhibiting `slope invariance' of the $T$-linear resistivity between physically distinct regimes as the temperature crosses the Hubbard-$U$ scale (with hopping $t \ll U$), providing a high temperature analogue of the phenomenon illustrated in Fig.~\ref{fig:martin}.

\section{Implications of a bound on dissipation}

In this closing section we will offer some thoughts on future work and argue that bounds, such as a possible Planckian bound, have a double role. Firstly, a useful bound expresses a limitation on physical processes, providing insight where a first principles approach to a given system may not be possible or available. Secondly, a bound can provide a hint of deeper understanding that may not yet have been fully developed.  
Both of these dimensions are illustrated with the example of
the Carnot bound on the efficiency of a heat engine in terms of the maximal accessible temperature difference. That bound contains a core truth about the working of engines that transcends any particular engineering design. It is rooted in the laws of thermodynamics, one of the most remarkable theories in the history of physics. Carnot and his contemporaries were driven by engineering, little knowing that the laws that they deduced from careful observations of the `classical' world would end up as a governing framework for many-body quantum mechanics.

We have described how the dissipative dynamics of several families of unconventional metals appear to be governed by a Planckian timescale. In many cases there is neither
a unique compelling candidate for a mode causing the $T$-linear transport nor a theoretical framework for computing the effects of the underlying scattering in a way that is both controlled and realistic. We have deliberately aired the complications and caveats associated with microscopic modelling and, indeed, the interpretation of experimental data.  We did so partly to give as complete account of the difficulties and subtleties as we can, but also to demonstrate what might be the central point of the whole field: These difficulties make the simplicity of the observed Planckian timescale all the more remarkable. In our opinion this fact strengthens the possibility that the ubiquity of the Planckian timescale reflects a quantum-mechanical limitation on the rate of dissipation.

The simplest point that we have made is that inelastic interactions renormalize the quasiparticle mass in tandem with scattering the quasiparticles. The Kramers-Kronig relations tie these two effects together and naturally lead to a Planckian bound on the physical quasiparticle lifetime in temperature regimes that have been probed experimentally. The bound is saturated in the limit of strong interactions.

Although the statements in the previous paragraph are couched in a quasiparticle language, they may reflect a more general physical principle. We have suggested that (some of) the observed Planckian transport lifetimes are a manifestation of an underlying timescale for many-body equilibration. The equilibration time and length are fundamental quantities for nonzero temperature many-body dynamics but have rarely been directly probed in condensed matter experiments. We have given arguments suggesting that neither of these quantities can become arbitrarily small. This is not a question of the validity of any given description, but of the consistency of the thermal state itself. These quantities are therefore promising starting points for formulating bounds on dynamics, and we discussed some routes towards achieving this. It would also be desirable to directly measure the equilibration time and length, possibly using ultrafast spectroscopy \cite{doi:10.1063/PT.3.1717}, or by characterizing the nature of fluctuations in the thermal state (taking inspiration, perhaps, from experiments on ultra-cold atomic systems \cite{Gross995} such as \cite{Gring1318, Trotzky2012,Kaufman794,PhysRevLett.117.170401,PhysRevX.8.021030}), or from measurements of spatially resolved transport (cf.~\cite{huang2021fingerprints}). For low temperature Planckian materials, these time- and length- scales may not be prohibitively short.

Suppose that there does exist, at least under some circumstances, a Planckian bound on quantum dynamics. Its existence will not in itself explain why systems are widely observed to saturate it. We highlighted the consequences of a bounded entropy; a possibly interesting future direction is that transport can often be formulated as a variational principle in which the dynamics extremizes the {\it rate} of entropy production, subject to certain constraints. See e.g.~\cite{zimanvar} for Boltzmann transport and \cite{Lucas11344} for inhomogeneous hydrodynamics. This may suggest that dissipative timescales will push up against a fundamental bound once all other bottlenecks, such as weak interactions between quasiparticles, are removed.

On the other hand, if a quasiparticle description is admissible then a different strategy may be more urgent. One should identify simple, realistic and material-specific scattering mechanisms for $T$-linear transport. The two case studies of quasiparticle Planckian transport that we have presented --- phonons and the marginal Fermi liquid --- may perhaps be a source of inspiration in this endeavour. One would also like to identify quasiparticle frameworks that allow different scattering mechanisms to exchange dominance without producing features in the resistivity. Needless to say, many microscopic quasiparticle scattering mechanisms have been suggested over the years but we believe it is fair to conclude that this problem has not been solved in general. At the same time, simple quantum-tuned phase diagrams like that shown in Fig.~\ref{fig:MFL} have been observed in an increasingly wide range of materials \cite{sachdevkeimer}.  Explaining these phase diagrams in detail, taking advantage of simultaneous transport and thermodynamic data, and using de Haas-van Alphen oscillation data to carefully account for the multi-band nature of many of these systems, is likely to provide strong constraints on theories of the underlying scattering. In this review we have not touched upon the rich phenomenology of magnetotransport \cite{Hayes2016, Sarkareaav6753, Giraldo-Gallo479, PhysRevResearch.1.023011, Nakajima2020, ayres2020incoherent, PhysRevX.10.041062, PhysRevResearch.2.033367} and low temperature thermal \cite{PhysRevX.8.041010} and thermoelectric \cite{PhysRevB.103.155102,doi:10.7566/JPSJ.90.053702} transport that has also been demonstrated experimentally in Planckian systems. These will likely also provide fruitful hunting ground for future theories.

In summary, we believe that a Planckian bound of some kind plausibly exists. Theoretically, establishing such a bound may entail the development of a quantum many-body `bootstrap' program, in the spirit of the highly successful use of bounds as organizing principles in critical systems \cite{Poland2016}.
Experimentally, there is a need for more spectroscopic data, particularly from the de Haas-van Alphen effect, optical conductivity and angle-resolved photoemission spectroscopy, on candidate Planckian systems.  We hope that this article will stimulate further research into this fascinating issue.

\section*{Acknowledgements}

It is a pleasure to acknowledge helpful discussions and comments on an earlier draft from Erez Berg, Andrey Chubukov, Luca Delacr\'etaz, Paolo Glorioso, Xizhi Han, Steve Kivelson, Andrew Lucas, Igor Markovi\'c, Chaitanya Murthy, Akshat Pandey, Brad Ramshaw, Jacob Ruf and Jan Zaanen, and Suzanna Stemmer for bringing Ref. \cite{DEVILLERS19841019} to our attention. We also wish to thank Elector Johann Georg III of Saxony, creator of the Grosser Garten in Dresden, for providing a beautiful and pandemic-proof setting in which this article was discussed during many long, socially-distanced walks. The work of S.A.H.~was partially supported by DOE award DE-SC0018134, by Simons Investigator Award \# 620869
and by STFC consolidated grant ST/T000694/1. S.A.H.~acknowledges the hospitality of the Max Planck Institute CPfS while this work carried out. A.P.M.~acknowledges the support of the Max Planck Society. The research environment in Dresden benefits from the Deutsche Forschungsgemeinschaft (DFG) Excellence Cluster ct.qmat, project ID 390858490.

\providecommand{\href}[2]{#2}\begingroup\raggedright\endgroup

\appendix

\newpage

\section{Diffusive timescales and diffusion bounds}
\label{sec:diff}

In \S\ref{sec:eq} we considered equilibration from the perspective of fast locally thermalizing processes. The subsequent diffusive evolution is more universal, yet also contains a trace of the underlying fast dynamics in the value of the diffusivity itself. The diffusivity, furthermore, is an unambiguously defined and measurable transport coefficient. This fact is especially important in strongly incoherent regimes where it may not be possible to define an effective mass and hence extract a timescale from transport. For this reason bounds have also been proposed on the diffusivity directly \cite{Kovtun:2004de, Hartnoll:2014lpa, Hartman:2017hhp}, as we discuss in this Appendix.

We can first explain why the diffusivity is naturally related to the underlying scales $\tau_\text{eq}$ and $\ell_\text{eq}$.
Inhomogeneities in the energy and charge density with small wavenumber $k$ necessarily decay slowly, because the homogeneous $k \to 0$ quantities (total energy and charge) cannot decay at all. Within linear response, i.e. for inhomogeneities with small amplitude, and focusing on charge dynamics\footnote{This is the charge dynamics prior to incorporating long-range electromagnetic interactions. Long-range Coulomb interactions screen charge fluctuations, that therefore decay exponentially quickly in time rather than diffuse. However, even in the presence of such Coulomb interactions it remains the case that the measured conductivity is given by $\sigma = \chi D$, with both the charge compressibility $\chi$ and diffusivity $D$ being intrinsic properties of the electronic system prior to the inclusion of long-range interactions. Different experimental probes will measure either the screened or unscreened Green's functions for the charge density. Screening complicates the direct measurement of the charge diffusivity in a metal, although this has recently been achieved in magic-angle twisted bilayer graphene \cite{Park2021}.} for concreteness, this slowness is expressed as an expansion in gradients
\be\label{eq:diffexpand}
\frac{\pa n}{\pa t} = D_2 \nabla^2 n + D_4 \nabla^4 n + \cdots  \,.
\ee
Here $D_2 = D$ is just the usual diffusivity. In order for (\ref{eq:diffexpand}) to make sense there must be some characteristic lengthscale above which it is possible to neglect the higher derivative terms. This is the thermalization length $\ell_\text{eq}$. Higher derivative corrections to the diffusion equation are suppressed at small wavenumber by $\ell_\text{eq} \nabla \ll 1$. This can be made explicit by writing $D_{2n} = \ell^{2n}_\text{eq}/\tau_{2n}$, where the timescales $\tau_{2n}$ are all comparable. In particular, the diffusivity $D = \ell_\text{eq}^2/\tau_2$.
These expressions, however, are not immediately useful because the timescales that appear will typically not be independent of $\ell_\text{eq}$. For example, $\ell_\text{eq}$ might be the quasiparticle mean free path and $\tau_{2n}$ of order the quasiparticle lifetime $\tau_\text{qp}$. These are related by the quasiparticle velocity $v_\text{qp}$. It is more instructive in this case to write the diffusivity as $D \sim v_\text{qp}^2 \tau_\text{qp}$.

Even without quasiparticles, the lengthscale in the diffusivity can be eliminated in favor of a velocity. Building on this perspective, it was argued in \cite{Hartman:2017hhp, Lucas:2017ibu, Han:2018hlj} that the diffusivity can always be {\it upper} bounded by writing it in terms of a certain many-body velocity $v$ and the equilibration time\footnote{If momentum is close to being conserved — for example because umklapp scattering is weak — then the equilibration time that controls the validity of the diffusive description of charge transport, and hence enters the bound (\ref{eq:upper}), is not the local equilibration time but the timescale at which diffusion crosses over to linearly dispersing sound propagation \cite{Hartman:2017hhp}. As momentum becomes increasingly conserved, both $D$ and this $\tau_\text{eq}$ diverge in (\ref{eq:upper}).}:
\be\label{eq:upper}
D \lesssim v^2 \tau_\text{eq} \,.
\ee
Introducing a velocity is well-motivated physically because in many cases the underlying microscopic dynamics proceeds ballistically. In particular, the ballistic growth of operators in space, discussed in \S\ref{sec:lyapunov}, leads to a microscopic `light-cone' causality constraint. Requiring that it should not be possible to diffuse outside the light-cone at early times leads to (\ref{eq:upper}). This logic is especially clear at very high temperatures where the relevant velocity is $v = v_\text{LR}$, the Lieb-Robinson velocity \cite{cmp/1103858407, hastings2010locality}, which sets an absolute speed limit to dynamics in local Hamiltonians. This is a physical velocity at high temperatures, see \cite{Brown379} for a measurement of diffusion at high temperatures with $D \sim v_\text{LR}^2 \tau_\text{eq}$. Away from high temperatures the relevant velocity is conceptually close to the butterfly velocity $v_B$ (discussed in \S\ref{sec:lyapunov}) but is not identical --- this point is discussed in \cite{Hartman:2017hhp, Lucas:2017ibu} and explored in holographic models in \cite{10.21468/SciPostPhys.9.1.007, Arean:2020eus, Wu:2021mkk, jeong2021breakdown}.

It has also been proposed, in the spirit of Planckian bounds, that the diffusivity obeys a {\it lower} bound $D \gtrsim v^2 \tau_\text{Pl}$ \cite{Hartnoll:2014lpa}. Unlike with the upper bound discussed above, a many-body velocity $v$ that could make this lower bound work in general has not been identified. For example, as we discussed extensively in the main text, in cases where quasiparticles undergo elastic scattering, the quasiparticle lifetime $\tau_\text{qp}$ can be much shorter than the Planckian time. In order for the diffusivity $D \sim v_\text{qp}^2 \tau_\text{qp}$ to obey the bound in such cases, it will be necessary that $v \ll v_\text{qp}$. A small $v$ will also be necessary 
close to localized phases, in order to avoid potential counterexamples such as charge transport in large $U$ Hubbard-like models \cite{PhysRevB.91.075124}.

Challenges for lower bounds on the diffusivity also arise in systems with a large number of degrees of freedom. If all of these degrees of freedom contribute to thermodynamics while only a few contribute to transport, then the diffusivity is very small. This fact is known to lead to violations of the $\eta/s$ bound, as mentioned in the original paper on the subject \cite{Kovtun:2004de}. In \cite{wu2021classical} a model was constructed in which a large number $N$ of incoherent optical phonons dominate the specific heat $c$ while only a single acoustic mode, with speed $v_s$ and lifetime $\tau_\text{ph}$, contributes significantly to the thermal conductivity $\kappa$. In this model the thermal diffusivity $D_\text{th} = \kappa/c \sim \frac{1}{N} v_s^2 \tau_\text{ph}$. Even if $\tau_\text{ph}$ is much larger than $\tau_\text{Pl}$, the diffusivity can be driven small by taking $N$ large. If such situations can be incorporated into a lower bound on diffusivity, it will again likely require a small velocity $v \ll v_s$.

Finally, note that the upper bound (\ref{eq:upper}) on the diffusivity can also be read as a lower bound on the equilibration time. This perspective is useful if something is known about the diffusivity, and was used in \cite{Delacretaz:2021ufg} to establish a stronger-than-Planckian lower bound on $\tau_\text{eq}$ in certain slowly-thermalizing two dimensional systems.

\section{Arguments for constraints on thermalization}

\subsection{Uncertainty principle constraint}
\label{sec:uncertain}

Here we give the details of the many-body uncertainty principle argument leading to (\ref{eq:tJ}) in the main text.
Consider a region of linear size $\ell_\text{eq}$ and recall from \S\ref{sec:eq} that such a region is able to thermalize itself on the timescale $\tau_\text{eq}$. The largest spread of energies that is available in the region is $\Delta E \lesssim J (\ell_\text{eq}/a)^d$. Here $a$ is the lattice spacing, $d$ the space dimension and $J$ the maximal energy range available at a single site. This argument requires a bounded on-site Hilbert space, excluding phonons. The uncertainty principle therefore implies that $J (\ell_\text{eq}/a)^d \tau_\text{eq} \gtrsim \hbar$. Furthermore, the local equilibration time cannot be shorter than the fastest time over which information can traverse the region.
The maximal physical velocity in a lattice system with spatially local interactions is the Lieb-Robinson velocity $v_\text{LR} \sim Ja/\hbar$ \cite{cmp/1103858407, hastings2010locality}. Here we assumed that the maximal `hopping' energy is comparable to the range of on-site energies $J$. Using $\tau_\text{eq} \geq \ell_\text{eq}/v_\text{LR}$ to eliminate $\ell_\text{eq}$ in the uncertainty relation above leads to
\be\label{eq:micro}
\tau_\text{eq} \gtrsim \frac{\hbar}{J} \,.
\ee

As we discuss in the main text, it is likely that this argument can be improved at low temperatures, because the range of on-site energies that can participate in the dynamics close to a low temperature thermal state will likely be narrower than the microscopic value $J$. It may also be possible to improve the argument at low temperatures by using the equilibration velocity (see Appendix \ref{sec:diff}) to relate $\ell_\text{eq}$ and $\tau_\text{eq}$. This velocity will often be less than the microscopic Lieb-Robinson velocity. 

\subsection{Bounding thermal fluctuations}
\label{sec:fluc}

Here we give details of the arguments leading to (\ref{eq:nell}) and footnote \ref{foot:cft} in the main text. As stated in the main text, the locally thermalized regions of finite size $\ell_\text{eq}$ will experience thermal fluctuations. The consequences of large fluctuations become especially sharp for operators that are bounded from below, such as the energy and particle number densities. If the variance in these quantities is larger than the distance of the thermal expectation value from the allowed minimum, then the microscopic positivity constraint is violated in a certain superposition of thermal microstates \cite{Delacretaz:2018cfk,dela}.

The energy in a thermalized region is $E = \epsilon \ell_\text{eq}^d$, with energy density $\epsilon$
normalized such that $\epsilon = 0$ in the ground state. The variance of the energy in the region is given by $\text{var}(E) = k_B T^2 \, c \ell_{eq}^d$, with $c$ the specific heat. Therefore imposing $\text{var}(E) \lesssim E^2$ gives
\be\label{eq:l1}
\ell_\text{eq}^d \gtrsim \frac{c}{k_B} \frac{(k_B T)^2}{\epsilon^2} \,.
\ee
Similarly, the variance for the particle number $N= n \ell_\text{eq}^d$ in the region is $\text{var}(N) = k_B T \chi \ell_\text{eq}^d$, with $\chi = dn/d\mu$ the charge compressibility. Again, the particle number is to be normalized such that $n=0$ is the lower bound. Bounding the variance in this case requires
\be\label{eq:l2}
\ell_\text{eq}^d \gtrsim \frac{k_B T \chi}{n^2} \,. \ee

Both (\ref{eq:l1}) and (\ref{eq:l2}) lower bound the equilibration time because $\tau_\text{eq} \geq \ell_\text{eq}/v_\text{LR}$ with $v_\text{LR}$ the maximal velocity discussed in the previous section.
For degenerate electrons, these lower bounds are weaker than Planckian bounds because $n, \epsilon$ and $\chi$ are temperature-independent while $c \sim T$. The right hand sides of the bounds therefore become small at low temperatures. However, as stated in footnote \ref{foot:cft} in the main text, a Planckian bound is obtained from (\ref{eq:l1}) in regimes where the thermodynamics is dominated by conformal field theory behavior (for example, close to a bosonic quantum critical point). Then $\epsilon \sim T^{d+1}$ and $c \sim T^d$ and the bound becomes $\tau_\text{eq} \gtrsim \hbar/(k_B T)$. Here we assumed that the temperature-independent scales making up the units in $\epsilon,c$ and $v_\text{LR}$ are comparable. In \cite{Delacretaz:2018cfk} this Planckian bound is obtained from within a field theory analysis, where positivity of energy is more subtle. 

For fermions around and above the degeneracy temperature $T_F$, the compressibility $k_B T \chi \sim n$ and hence the
bound (\ref{eq:l2}) becomes $n \ell_\text{eq}^d \gtrsim 1$. Because $\ell_\text{eq}$ is expected to decrease with increasing temperature in a metal, this bound likely also holds at lower temperatures where the fermions are degenerate. While this represents a strictly unjustified logical leap, the bound is also, as we noted in the main text, the intuitive statement that for a region to be able to self-thermalize it should contain more than one particle.
This is the statement of (\ref{eq:nell}) in the main text.

\section{The Kadowaki-Woods ratio}
\label{sec:KW}

In the main text we have discussed mass renormalization in detail. An important additional perspective on the role of mass renormalization in transport comes from the Kadowaki-Woods ratio. In a Fermi liquid the resistivity $\rho = A T^2$ and the specific heat coefficient $\gamma = c/T$ is temperature-independent. The Kadowaki-Woods ratio  $A/\gamma^2$ was observed to be roughly the same across different heavy fermion materials even while strong mass renormalization meant that $A$ and $\gamma$ separately varied over orders of magnitude \cite{KADOWAKI1986507}. We recalled in (\ref{eq:gamma}) that  $\gamma  \propto 1/Z \propto m_\star$. The observation therefore implies that the resistivity coefficient $A \propto m_\star^2$. As we have discussed in the main text, in materials with multiple sheets it is prudent to obtain $m_\star$ from quantum oscillations rather than the specific heat; the ratio $A/\gamma^2$ need not stay constant for mass divergences associated to a heavy sheet that does not contribute to transport \cite{Mousatov2852}.

We will firstly summarize the argument in e.g.~\cite{MIYAKE19891149, Jacko2009} showing that, with certain assumptions, the scaling $A \propto m_\star^2$ in a Fermi liquid follows from the Kramers-Kronig relation. We recalled in \S\ref{sec:drude2} that in a Fermi liquid
\be\label{eq:SFL}
\Sigma''(\omega) = - \a^2 \left[\omega^2 + (\pi k_B T/\hbar)^2 \right] \,,
\ee
for frequencies $0 \leq |\omega| < \omega_\text{c}$.
If the strength of scattering is `order one' at the cutoff $\omega_\text{c}$ then one expects $\a^2 \propto 1/\omega_\text{c}^2$. We will assume this scaling in the following.
Using (\ref{eq:SFL}) in the Kramers-Kronig relation (\ref{eq:overZ}) gives the mass renormalization
\be\label{eq:DeltamFL}
\frac{m_\star}{m} =1 +  \frac{2 \a^2 \omega_\text{c}}{\pi} \propto \frac{1}{\omega_\text{c}} \,.
\ee
In the second step we have assumed that the mass renormalization dominates the bare term. The relation $\omega_\text{c} \propto 1/m_\star$ implies that the cutoff is proportional to the renormalized Fermi energy $E_{F\star}$. Finally, we must also assume that $|\Sigma''(\omega)|$ does not continue increasing beyond the cutoff frequency $\omega_\text{c}$. With that assumption, while the contribution from high frequencies $\omega > \omega_\text{c}$ may alter the coefficient of the correction in (\ref{eq:DeltamFL}) they will not change its dependence on $\omega_\text{c}$. Thus we conclude that $\alpha^2 \propto 1/\omega_\text{c}^2 \propto m_\star^2$.

Recall from the discussion below (\ref{eq:dc}) in the main text that the resistivity can be written in terms of unrenormalized quantities as $\rho = m |\Sigma''(0)|/(n e^2)$. Using the self-energy (\ref{eq:SFL}) gives
\be
\rho  \propto \a^2 T^2 \propto m_\star^2 T^2 \,,
\ee
which is the desired result. Writing $\rho \propto m_\star/\tau$ in terms of the physical mass and scattering rate gives
\be\label{eq:tauFL}
\frac{1}{\tau} \propto m_\star T^2 \propto \frac{T}{E_{F\star}} \times \frac{k_B T}{\hbar} \ll \frac{1}{\tau_\text{Pl}} \,.
\ee
This expression for the quasiparticle lifetime agrees with the result obtained from a simple-minded Fermi golden rule scattering computation that uses the renormalized mass to obtain the scattering phase space and an electronic scattering strength that is order one in units of the renormalized mass. It serves as a nice illustration of the point made in \S \ref{sec:transportime}: to extract the physical lifetime in a Drude-style analysis, the renormalized mass should be used.

This simple-minded perspective can be legimitized from the viewpoint of the Wilsonian renormalization group, as we now explain. The computation of the decay rate only involves modes within an energy width $T$ of the Fermi surface. One may obtain an effective Hamiltonian for these low energy modes by integrating out all modes at energies above $T$. If the temperature $T \ll \omega_\text{c}$, this integration over modes with frequencies from $T$ to $\omega_\text{c}$ accounts for essentially the entire mass renormalization 
(\ref{eq:DeltamFL}). The effective low energy Hamiltonian therefore contains the renormalized mass $m_\star$, which should then be used both in Fermi's golden rule and in the Drude formula. If one finally assumes that the low energy Hamiltonian is `natural', that is, all dimensionless coupling constants in the effective theory are order one, then one obtains (\ref{eq:tauFL}) as well as $\rho \propto m_\star^2 T^2$.

Let us now consider the case where the Fermi liquid
arises below a collapsing `quantum critical fan', as in Fig.~\ref{fig:masses} in the main text. Away from the quantum critical point, the critical modes are gapped at an energy scale $\Delta$. In the low temperature Fermi liquid regime $T \ll \Delta$ the critical modes can be integrated out along with the high energy electronic modes to give an effective Hamiltonian for physics at energy scale $T$, where all high energy dynamics has been subsumed into the renormalized mass $m_\star$. The result 
$\rho \propto m_\star^2 T^2$ will then follow as previously. Consistent with this conclusion,
measurements in the iron-based superconductor BaFe$_2$(As$_{1-x}$P$_x$)$_2$ \cite{Analytis2014, PhysRevLett.110.257002} and in the heavy fermion material CeRhIn$_5$ \cite{jpsj, doi:10.1143/JPSJ.74.1103} suggest that the coefficient of the Fermi liquid $T^2$ resistivity does indeed suitably track the divergence of the effective mass, as measured by quantum oscillations, upon approaching the critical point at fixed low temperature.

\end{document}